\documentclass[aps,pra,twocolumn,showpacs,superscriptaddress,preprintnumbers,amsmath,amssymb,footinbib]{revtex4-1}

\usepackage{xcolor}
\usepackage[breaklinks=true,colorlinks,citecolor=blue,linkcolor=blue,urlcolor=blue]{hyperref}

\usepackage{chemformula} 
\usepackage[T1]{fontenc} 
\usepackage{amsmath}
\usepackage{amsfonts}
\usepackage{color}
\usepackage{float}
\usepackage{tabularray}
\usepackage{tabularx}
\usepackage{multirow}
\usepackage{booktabs}
\usepackage{atbegshi}
\usepackage{upgreek}
\DeclareMathSymbol{\shortminus}{\mathbin}{AMSa}{"39}

\usepackage{parskip}
\usepackage{xfp}
\newcommand\SupplementaryMaterials{%
  \xdef\presupfigures{\arabic{figure}}
  \xdef\presuptables{\arabic{table}}
  \xdef\presupsections{\arabic{section}}
  \xdef\presupsubsections{\arabic{subsection}}
  \renewcommand\thefigure{S\fpeval{\arabic{figure}-\presupfigures}}
  \renewcommand\thetable{S\fpeval{\arabic{table}-\presuptables}}
  \renewcommand\thesection{}
  \renewcommand\thesubsection{Note~\fpeval{\arabic{subsection}-\presupsubsections}}

}

\usepackage{array}
\newcolumntype{x}[1]{>{\centering\arraybackslash\hspace{0pt}}p{#1}}

\begin{document}

\title{A pre-time-zero spatiotemporal microscopy technique for the ultrasensitive determination of the thermal diffusivity of thin films}
\author{Sebin Varghese} \affiliation{Catalan Institute of Nanoscience and Nanotechnology (ICN2), BIST and CSIC, Campus UAB, 08193 Bellaterra (Barcelona), Spain}
\author{Jake Dudley Mehew} \affiliation{Catalan Institute of Nanoscience and Nanotechnology (ICN2), BIST and CSIC, Campus UAB, 08193 Bellaterra (Barcelona), Spain}
\author{Alexander Block} \affiliation{Catalan Institute of Nanoscience and Nanotechnology (ICN2), BIST and CSIC, Campus UAB, 08193 Bellaterra (Barcelona), Spain}
\author{David Saleta Reig} \affiliation{Catalan Institute of Nanoscience and Nanotechnology (ICN2), BIST and CSIC, Campus UAB, 08193 Bellaterra (Barcelona), Spain}
\author{Paweł Wo{\' z}niak}\affiliation{ICFO $-$ Institut de Ciències Fotòniques, Mediterranean Technology Park, Castelldefels (Barcelona) 08860, Spain}
\author{Roberta Farris} \affiliation{Catalan Institute of Nanoscience and Nanotechnology (ICN2), BIST and CSIC, Campus UAB, 08193 Bellaterra (Barcelona), Spain}
\author{Zeila Zanolli} \affiliation{Chemistry Department and ETSF, Debye Institute for Nanomaterials Science, Utrecht University, the Netherlands}
\author{Pablo Ordejón} \affiliation{Catalan Institute of Nanoscience and Nanotechnology (ICN2), BIST and CSIC, Campus UAB, 08193 Bellaterra (Barcelona), Spain}
\author{Matthieu J. Verstraete} \affiliation{Nanomat, Q-Mat, CESAM, and European Theoretical Spectroscopy Facility, Université de Liège, B-4000 Liège, Belgium}
\author{Niek F. van Hulst}\affiliation{ICFO $-$ Institut de Ciències Fotòniques, Mediterranean Technology Park, Castelldefels (Barcelona) 08860, Spain} \affiliation{ICREA, Pg. Lluís Companys 23, 08010 Barcelona, Spain}
\author{Klaas-Jan Tielrooij} \email[Correspondence to: ]{klaas.tielrooij@icn2.cat} \affiliation{Catalan Institute of Nanoscience and Nanotechnology (ICN2), BIST and CSIC, Campus UAB, 08193 Bellaterra (Barcelona), Spain}

\begin{abstract}
\vspace{0.5cm}
Diffusion is one of the most ubiquitous transport phenomena in nature. Experimentally, it can be tracked by following point spreading in space and time. Here, we introduce a spatiotemporal pump-probe microscopy technique that exploits the residual spatial temperature profile obtained through the transient reflectivity when probe pulses arrive before pump pulses. This corresponds to an effective pump-probe time delay of 13 ns, determined by the repetition rate of our laser system (76~MHz). This pre-time-zero technique enables probing the diffusion of long-lived excitations created by previous pump pulses with nanometer accuracy, and is particularly powerful for following in-plane heat diffusion in thin films. In contrast to existing techniques for quantifying thermal transport it does not require any material input parameters or strong heating. We demonstrate the direct determination of the thermal diffusivities of the layered materials MoSe$_2$ (0.18~cm$^2$/s), WSe$_2$ (0.20~cm$^2$/s), MoS$_2$ (0.35~cm$^2$/s), and WS$_2$ (0.59~cm$^2$/s). This technique paves the way for observing novel nanoscale thermal transport phenomena and tracking diffusion of a broad range of species.
\end{abstract}

\maketitle

\section{Introduction}

Pump-probe measurements, where a probe pulse interrogates a system at a variable time delay after the system has been excited by a pump pulse, are powerful and versatile techniques to obtain information on dynamical processes with high temporal resolution and with high sensitivity. In spatiotemporal pump-probe measurements \cite{Ruzicka2010, Gabriel2013,Kumar2014,Wan2015,Block2019,Sung2020,seitz2020} pump and probe pulses have a variable relative offset in space, in addition to their variable time delay. Typically, one of the beams is scanned over the other, fixed, beam, providing information on spatiotemporal transport phenomena, such as diffusion. Most commonly, one observes broadening of an initially pump-excited spot as a function of pump-probe time delay $\Delta t$, where the probe arrives later than the pump. The accuracy in resolving spatial broadening is determined by the signal-to-noise ratio, and can therefore be significantly beyond the diffraction limit. Recently, this super-resolution technique has gained increasing attention as an important tool for studying diffusion in semiconductors \cite{Ruzicka2010,Gabriel2013}, organic compounds \cite{Wan2015}, perovskites \cite{Sung2020,seitz2020}, metals \cite{Block2019}, two-dimensional materials \cite{Kumar2014,block2021}, and more. However, with current spatiotemporal pump-probe techniques it is challenging to resolve processes with low diffusivities.

One specific slow diffusion process is the phonon heat diffusion that governs the lattice thermal conductivity \cite{kittel2021}. Understanding thermal transport is extremely important, in particular when materials are used in technological applications: avoiding (local) overheating is crucial for proper device operation and prolonged device lifetimes in electronic, optical and optoelectronic devices and systems \cite{tong2011}. However, it is highly challenging for spatiotemporal techniques based on pump-probe measurements to resolve thermal diffusion, because of the limited amount of broadening. Silicon, for example, has a thermal diffusivity $D$ around 1 cm$^2$/s \cite{shanks1963}. If heat is deposited by a pump pulse in a Gaussian spot with a width $\sigma_{\rm \Delta t = 0}$, the two-dimensional broadening as a function of pump-probe delay time $ \Delta t$ is described by the diffusion equation: $\sigma^2_{\rm \Delta t > 0} = \sigma^2_{\rm \Delta t = 0} + 2D \Delta t$. This means that with an initial spot size of 1 $\upmu$m, after 100 ps the broadening due to thermal diffusion in silicon is just a few nanometers. Moreover, most materials have a lower diffusivity than silicon, and therefore show even less broadening. Alternative optical techniques to measure thermal transport typically obtain the thermal conductivity rather than the diffusivity, require accurate knowledge of several material and interface parameters, and/or need relatively strong heating of tens to hundreds of Kelvin. Moreover, the thermal diffusivities of even the most typical transition metal dichalcogenide (TMD) thin films --  MoSe$_2$, WSe$_2$, MoS$_2$, and WS$_2$ -- have not yet been obtained in a direct experimental way.

Here, we introduce pre-time-zero spatiotemporal microscopy as a technique that enables the direct determination of the diffusivity of long-lived excitations by examining the spatial profile at a small negative pump-probe delay time, where the probe arrives before the pump. In this configuration, the probe is sensitive to residual signal generated by previous pump pulses from the incident pulse train. For example, with a repetition rate $f_{\rm rep}$ of 76 MHz, the pre-time-zero signal corresponds to an effective time delay of $\Delta t \approx$ 13 ns. We demonstrate the strengths of this technique by quantifying the in-plane thermal diffusivity $D$ of free-standing thin films of the TMD material family. We use suspended films, such that phonon heat can accumulate due to the absence of a heat-sinking substrate. Furthermore, the temperature sensitivity reaches the sub-Kelvin level, when probing with a wavelength close to the exciton resonance, where the reflection is particularly sensitive to lattice temperature. We obtain the thermal diffusivity directly by comparing the experimental spatial profile to a simple model based on Fourier’s law of heat conduction, which does not require any material input parameters. The obtained experimental results are in agreement with our \textit{ab initio} simulations based on density functional theory (DFT) and with experimental values in the literature. We will discuss the advantages of this technique compared to existing optical techniques for the determination of the thermal diffusivity and conductivity. Furthermore, we also provide an outlook of the important studies this technique enables, in terms of (nanoscale) thermal transport, and beyond. 

\section*{RESULTS}

\subsection*{Concept and implementation of pre-time-zero spatiotemporal microscopy}

\begin{figure*}
  \includegraphics[width = 13cm]{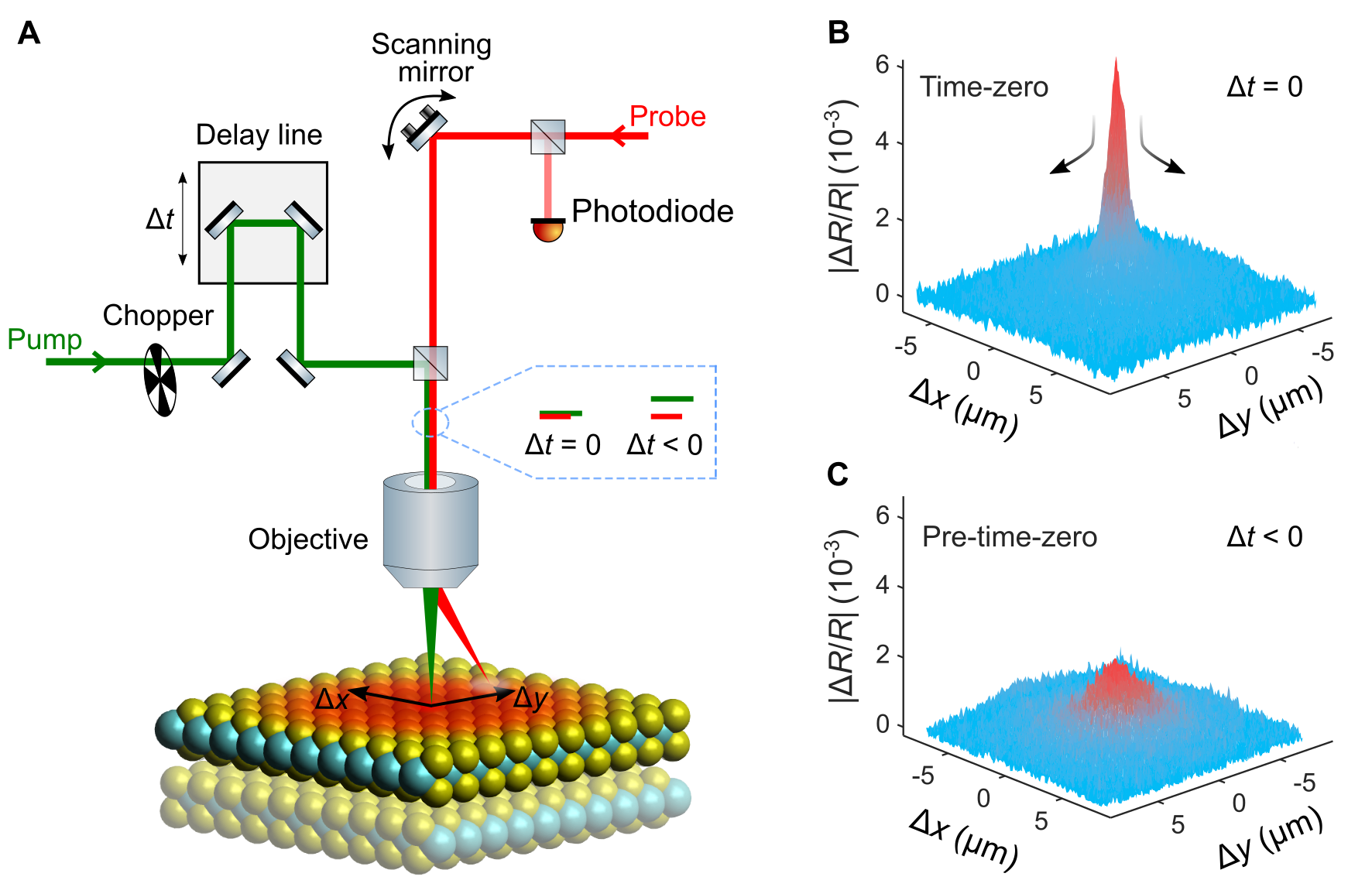}
  \caption{\textbf{Schematic representation of the spatiotemporal pump-probe microscopy setup.} \textbf{(A)}~An optical pump pulse is focused onto the sample, creating a local excitation. To monitor the diffusion of this excitation, a probe beam scans spatially over the sample around the pump spot. The modulated pump beam passes over a delay line that controls the time delay $\Delta t$ between the pulses. The probe beam position at the sample plane is controlled by a scanning mirror. Both beams are combined with a dichroic mirror and focused onto the sample through an objective. The reflected probe beam is detected using a silicon photodiode and demodulated at the modulation frequency, typically in the kHz range. A detailed schematic is shown in Supplementary Materials, Fig.~\ref{fig:SM_setup}. \textbf{(B)}~The transient reflectivity profile $\frac{\Delta R}{R}(\Delta x, \Delta y)$ at time delay $\Delta t$ = 0 (time-zero) shows a narrow profile that corresponds to the photoexcitation from the pump pulse. \textbf{(C)}~The profile measured at a small negative time delay $\Delta t <$ 0 (pre-time-zero) shows a broadened response due to diffusion of excitation created by previous pump pulses. Effectively this corresponds to a pump-probe delay that is the inverse of the laser repetition rate, typically in the nanosecond regime. }
  \label{fgr:one}
\end{figure*}

We explain the fundamental idea behind our pre-time-zero spatiotemporal technique, and the experimental setup, in Fig.~\ref{fgr:one}A. An optical pump pulse excites a small region of the sample. Initially, the spatial distribution of the excited species corresponds to the profile of the tightly focused pump beam. As the excited species diffuse, the profile broadens spatially. In the case that the diffusing excitation lives long enough for a new pump pulse to arrive before the system has returned to equilibrium, we can monitor the spatially broadened profile using a probe pulse at a small negative pump-probe time delay. This delay time thus corresponds to a delay that is roughly equal to the temporal spacing between subsequent pump pulses. This enables the study of spatial broadening occurring over a longer timescale than the conventional pump-probe delay time.

Whereas the concept is generally applicable to any long-lived excitations, we now focus on the specific case of quantifying thermal transport in thin films, in order to demonstrate the strengths of pre-time-zero spatiotemporal pump-probe microscopy. We exploit the notion that a pump-induced electronic excitation relaxes over time, typically transferring energy to the lattice subsystem, where phonons carry the energy as heat. We furthermore use the idea that heat changes the optical properties of a material, which we detect with our probe pulses. Since pump-probe measurements enable the observation of very small pump-induced changes in transmission or reflection, this technique can detect small amounts of heat, which we will quantify below. We fix the position of the pump beam, and  scan the probe beam in the sample plane, resolving the spatial profile of the transient reflectivity signal $\frac{\Delta R}{R}$ (detected using a silicon photodiode) as a function of spatial pump-probe offset $\Delta x$ and/or $\Delta y$. Here, transient reflectivity is $\frac{\Delta R}{R} = \frac{R_{\rm exc} - R}{R}$, where $R_{\rm exc}$ and $R$ are the reflection of the probe with and without the presence of the pump pulse, respectively. By scanning the position of the probe beam with respect to the pump beam with nanometer precision, we acquire accurate spatial transient reflectivity maps $\frac{\Delta R}{R}(\Delta x, \Delta y)$. We will show that the transient reflectivity profile represents the temperature profile $\Delta T(\Delta x, \Delta y)$, which enables us to extract the thermal diffusivity.

In our specific experimental setup (see Supplementary Materials, Fig.~\ref{fig:SM_setup}), a mode-locked laser ($f_{\rm rep} = 76$~MHz) generates pulses centered at $1030$~nm. Most of the laser output power is used to pump an optical parametric oscillator (OPO) which has a tunable signal output between $1320$ and $2000$~nm. The probe beam is either the second or the third harmonic of the signal output, depending on the sample under investigation. We direct the probe via a scanning mirror system (Optics in Motion OIM101) while the pump beam (515~nm, second harmonic of the fundamental laser source) goes onto a variable delay line (Newport DL225) and an optical chopper that regulates the pump modulation frequency $f_{\rm mod}$. Both beams combine via a dichroic mirror and focus to sub-micron spot sizes via a microscope objective lens (numerical aperture: 0.67).

Figure~\ref{fgr:one}B and \ref{fgr:one}C show exemplary transient reflectivity maps recorded at two pump-probe time delays $\Delta t$. At time-zero ($\Delta t$~=~0), we observe a narrow profile centered at the pump spot $\Delta x$~=~$\Delta y$~=~0, which is due to the electronic excitation induced by absorption of the pump pulse, see Fig.~\ref{fgr:one}B. The spatial profile looks very different when tuning to a small negative time delay $\Delta t < 0$, \textit{i.e.}\ pre-time-zero, typically a few picoseconds before time-zero. Here we observe a broader profile with a smaller amplitude, see Fig.~\ref{fgr:one}C. We attribute this signal to diffused phonon heat in the sample. The broader the pre-time-zero profile is, the larger we expect the diffusivity of the phonon heat to be, and this eventually allows us to obtain the diffusivity quantitatively.

\subsection*{Spatiotemporal pump-probe measurements on TMDs}

We apply our spatiotemporal technique to a prototypical thin-film system: transition metal dichalcogenide crystals. Like graphite, TMDs are layered crystals with strong in-plane bonds between atoms, while adjacent layers are weakly bound by van der Waals forces. They have attracted significant interest in recent years due to their many intriguing optical \cite{Gutierrez2013}, electrical \cite{neupane2017}, mechanical \cite{Babacic2021}, and thermal properties \cite{Jiang2017}. A thorough understanding of their thermal properties is crucial to enable technologies, including (opto)electronic, thermoelectric, medical, and thermal management applications. Whereas their thermal conductivities $\kappa$ have been obtained experimentally, their thermal diffusivities $D$ have not been determined in a direct way, even though the diffusivity arguably describes thermal transport in a more direct fashion.  We study free-standing films of MoSe$_2$, WSe$_2$, MoS$_2$ and WS$_2$, of thickness around 15~nm, corresponding to $\sim$20 layers, suspended over circular holes with a diameter of 15 $\upmu$m (Norcada, NTPR005D-C15). The sample fabrication, based on PDMS-assisted dry transfer of mechanically exfoliated TMDs (purchased from HQ graphene), is described in detail elsewhere \cite{Varghese_2021}. To facilitate efficient heat sinking outside the suspended area, we coated the substrate with 5/50~nm of Ti/Au. The measurements are performed in air (and at room temperature), as going to vacuum led to negligibly small differences (see Supplementary Materials, Fig.~\ref{fgr:SM_vacuum}).

\begin{figure*}[ht!]
  \includegraphics[width = 13cm]{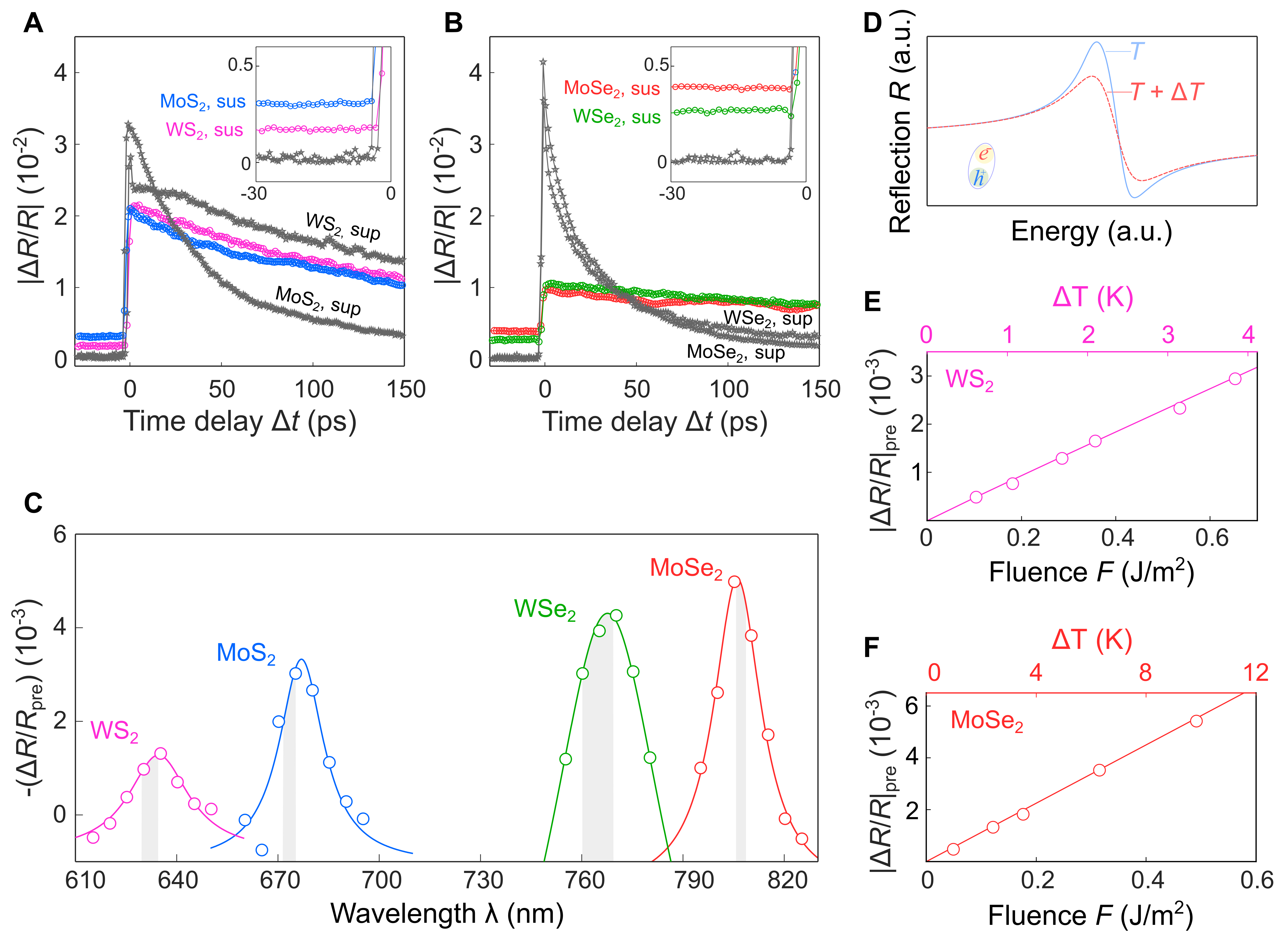}
  \caption{\textbf{Pre-time-zero spatiotemporal experiment.} Comparison of absolute transient reflectivity $|\frac{\Delta R}{R}|$ dynamics in suspended (sus) and supported (sup) regions of WS$_2$ and MoS$_2$~\textbf{(A)}, WSe$_2$ and MoSe$_2$~\textbf{(B)}. Insets show a zoomed in view of the signal before time zero. \textbf{(C)}~Pre-time-zero reflectivity signal ($| \frac{\Delta R}{R} |_{\rm \Delta t < 0}$) measured at different probe wavelengths with a constant incident pump fluence of ${\sim 0.3}$~J/m${^2}$. The largest response corresponds to the A exciton resonance of the materials. The solid line is a Lorentzian fit to the data. The shaded area represents typical A exciton peaks found in literature. \textbf{(D)}~Simulated reflection spectrum around an exciton resonance at two different temperatures. \textbf{(E, F)}~Magnitude of the pre-time-zero signal as a function of pump fluence (circles,~bottom axis) and temperature rise (solid line, top axis), for WS$_2$~\textbf{(E)} and MoSe$_2$~\textbf{(F)} extracted from the Lorentz oscillator model. For \textbf{(C, E, and F)}~we average $\sim$50 data points at different negative time delays from an interval of a few picoseconds. The error bars are smaller than the marker size (68\% confidence interval) and the exponent of a power law fit to these data is 0.98 (\textbf{E}) and 0.99 (\textbf{F}), indicating that the relation is very close to linear.}
  \label{fgr:two}
\end{figure*}

Figure~\ref{fgr:two}A and \ref{fgr:two}B show the transient reflectivity signal on suspended and supported regions of the flakes, as a function of pump-probe delay time. In both regions, the signal peaks at $\Delta t$~=~0~ps. This is associated with the photoexcitation of charge carriers and sub-picosecond exciton formation \cite{Kumar2014}. We first examine the substrate-supported region, where the signal decays over picosecond timescales and can be understood as a decrease in exciton population \cite{Cui2014}. The pump-probe signal decays to a value very close to zero during the time window provided by our delay line (150 ps), and is basically zero after $\sim$13 ns, as observed in the pre-time-zero signal. This means that the system is in equilibrium before the next pump pulse arrives. In the suspended region, in contrast, there is a clear remnant signal before time zero, see Fig.~\ref{fgr:two}A and \ref{fgr:two}B. We ascribe this signal to phonon heat that has accumulated in the sample. This does not occur in the supported regions due to efficient out-of-plane heat sinking to the gold-coated substrate.

In order to confirm that the pre-time-zero signal indeed corresponds to a phonon-heat induced change in the reflection of our probe, we measure the signal as a function of probe wavelength (see Fig.~\ref{fgr:two}C). We observe a clear peak at a wavelength of $635$~nm, $675$~nm, $770$~nm, and $805$~nm which corresponds to the A-exciton resonance of WS$_2$, MoS$_2$, WSe$_2$, and MoSe$_2$, respectively\cite{dong2015,niu2018,zhao2013}. We describe the experimental data with a Lorentzian function and extract linewidths of $21$~nm (WS$_2$), $16$~nm (MoS$_2$), $44$~nm (WSe$_2$), and $17$~nm (MoSe$_2$), similar to the literature reports of exciton linewidths \cite{mohamed2017}. It is known that these linewidths are temperature dependent \cite{Selig2016}, which provides the physical connection between the measured optical response and phonon heat in the sample: Figure~\ref{fgr:two}D shows a representative reflection profile obtained from the Lorentz oscillator model (see Supplementary Materials,~\ref{Note1}) at two different phonon temperatures. With an increased lattice temperature, the exciton linewidth increases, and this leads to a decrease in reflection, which is indeed what we observe.

We examine this in more detail by measuring the transient reflectivity signal as a function of pump fluence $F$ (see Fig.~\ref{fgr:two}E and~\ref{fgr:two}F) in WS$_2$ and MoSe$_2$. The first thing we notice is the linear dependence of the transient reflectivity $\frac{\Delta R}{R}$ on fluence (also see Supplementary Materials, Fig.~\ref{fgr:SM_fluence}). Importantly, this justifies the use of the transient reflectivity profile $\frac{\Delta R}{R} (\Delta x, \Delta y)$ as a representative of the temperature profile $\Delta T (\Delta x, \Delta y)$, from which we will extract the diffusivity. Secondly, we extract the temperature increase $\Delta T$ from the observed transient reflectivity, and compare this to the expected $\Delta T$ based on pulse energy and heat capacity. For this, we use the complex dielectric function of WS$_2$ and MoSe$_2$ according to a Lorentz oscillator model of the exciton resonance (see Supplementary Materials,~\ref{Note1}), and the temperature-dependent exciton linewidth from Ref.\ \cite{Selig2016}, finding values below 10 K up to the largest applied fluence (see Fig.~\ref{fgr:two}E and \ref{fgr:two}F). We compare this extracted $\Delta T$ to the peak temperature increase using the relation $\Delta T_{\rm peak} = (A\cdot F)/(h \cdot C_{\rm v})$, where $A$ is the absorption (see Supplementary Materials, Fig.~\ref{fgr:SM_absorption}), $h$ is the thickness, and $C_{\rm v}$ is the heat capacity, finding good agreement. A fluence of $0.4$~J/m$^{2}$, for example, in WS$_2$ (MoSe$_2$) corresponds to a $\Delta T$ $\sim 2$~K ($\sim 8$~K) from the Lorentz model (see Fig.~\ref{fgr:two}E and \ref{fgr:two}F) and a peak temperature rise of $\Delta T_{\rm peak} \sim 3$~K ($\sim 5$~K), using $h = 17.2$~nm (13.8~nm), and $C_{\rm v}$ from literature (WS$_2$ from Ref.\cite{ohare1984} and MoSe$_2$ from Ref.\cite{blinder1993}). Importantly, these results show that the technique is sensitive to reflection changes caused by heating by less than a Kelvin. 

\begin{figure*}
  \includegraphics[width = 14cm]{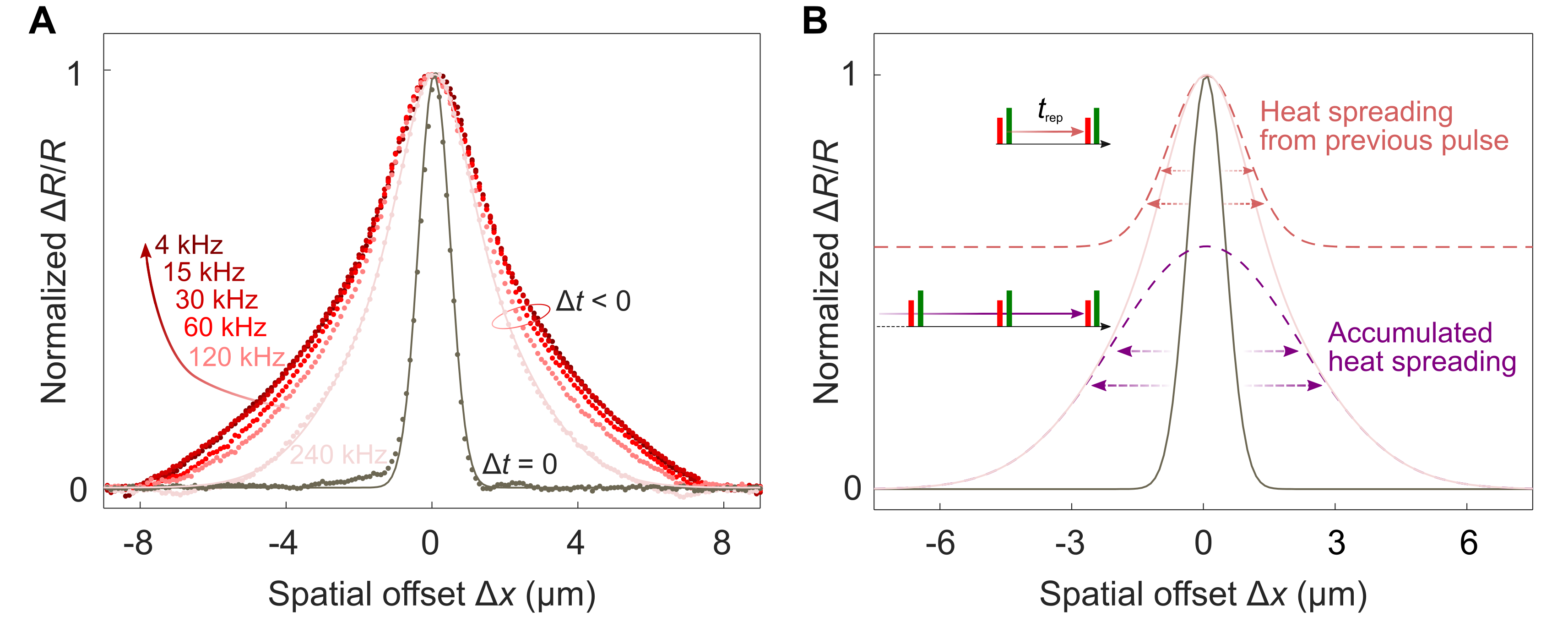}
  \caption{\textbf{Spatial profiles at varying modulation frequencies for MoSe$_2$.} \textbf{(A)}~ Spatial profiles $\Delta t < 0$ for different modulation frequencies. For low modulation frequencies (4 -- 30~kHz), the spatial profiles are similar, indicating temperature saturation due to steady state formation. The solid line for $\Delta t < 0$ profile (240~kHz) is the description of the data by the sum of two Gaussians. The narrow profile represents the corresponding $\Delta t = 0$ profile (grey data points) and is described by a single Gaussian function. \textbf{(B)}~Description of the spatial profile with 240~kHz modulation at $\Delta t = 0$ (grey line, single Gaussian) and $\Delta t < 0$ (light red line, sum of two Gaussians). The narrow dashed Gaussian (offset vertically) represents the resulting profile due to heat spreading from the previous pump pulse. The broad Gaussian is the resulting profile due to accumulative heat spreading from all the previous pump pulses. By analysing the width of the $\Delta t = 0$ profile and the width of the narrow Gaussian, we obtain a $D$ of 0.21~cm$^2$/s. All profiles are normalized to the peak of the fit.} 
  \label{fgr:three}
\end{figure*}

Having established that our probe is highly sensitive to phonon heat via the exciton linewidth, we now examine the $\Delta t < 0$ spatial profiles at different pump modulation frequencies ($f_{\rm mod}$) between 4 and 240 kHz, as controlled by an optical chopper or an electro-optic modulator. We position the pump beam in the center of the suspended region of the sample and scan the probe beam with respect to the pump spot at $\Delta t = 0$ and at a small negative time delay ($\Delta t < 0$), with a pump fluence of $\sim$0.3~J/m$^{2}$. Figure~\ref{fgr:three}A shows the normalized $\frac{\Delta R}{R}$ spatial profiles acquired at $\Delta t = 0$ and $\Delta t < 0$ for MoSe$_2$. For the $\Delta t = 0$ profile, we subtracted the $\Delta t < 0$ background signal (see Supplementary Materials, Fig.~\ref{fgr:SM_background}), and find that it is accurately described using a Gaussian function (Fig.~\ref{fgr:three}A), which corresponds to a convolution of our pump and probe spot sizes (see Supplementary Materials, Fig.~\ref{fgr:SM_knifeedge} and ~\ref{Note2}). We observe that the $\Delta t < 0$ spatial profiles broaden with decreasing $f_{\rm mod}$ and tend to saturate for the lowest modulation frequencies, see Fig.~\ref{fgr:three}A. This saturation is due to the formation of a steady-state profile and is understood as follows: The rate with which heat that is added to the material by absorbed light from the pump pulses is equal to the rate with which heat diffuses to the edge of the suspended material and heat-sinks into the gold-coated substrate. These spatial profiles can be described by two Gaussians: The narrow component corresponds to the diffused heat from the previous pump pulse and is therefore associated with a timescale of 1/$f_{\rm rep}$ = 13 ns. The broader component corresponds to accumulated heat spreading from all the previous pump pulses and is therefore associated with a timescale of 1/$f_{\rm mod}$ = 4 -- 250 $\upmu$s. For lower modulation frequencies, corresponding to longer excitation with pump pulses, there is more accumulated heat, making the profile broader. We analyse the spatial profile taken at 240~kHz and use its narrow component to estimate the thermal diffusivity $D$ of the sample. This is done by comparing the Gaussian width ($\sigma_{\Delta t = 0}$) of the spatial profile at $\Delta t = 0$ and the Gaussian width ($\sigma_{\rm \Delta t < 0, narrow}$) of the narrow part of the $\Delta t < 0$ profile. The diffusivity is then given by $D = [\sigma^2_{\rm \Delta t < 0, narrow} - \sigma^2 _{\Delta t = 0)}]/2 \Delta t$. Using $\Delta t$ = 13 ns, we obtain a $D$ of $0.21$~cm$^2$⁄s, see Fig.~\ref{fgr:three}B. We will show below that this is a very reasonable estimate.

\subsection*{Simulations of the pre-time-zero spatiotemporal technique}

\begin{figure*}
  \includegraphics[width = 13.5cm]{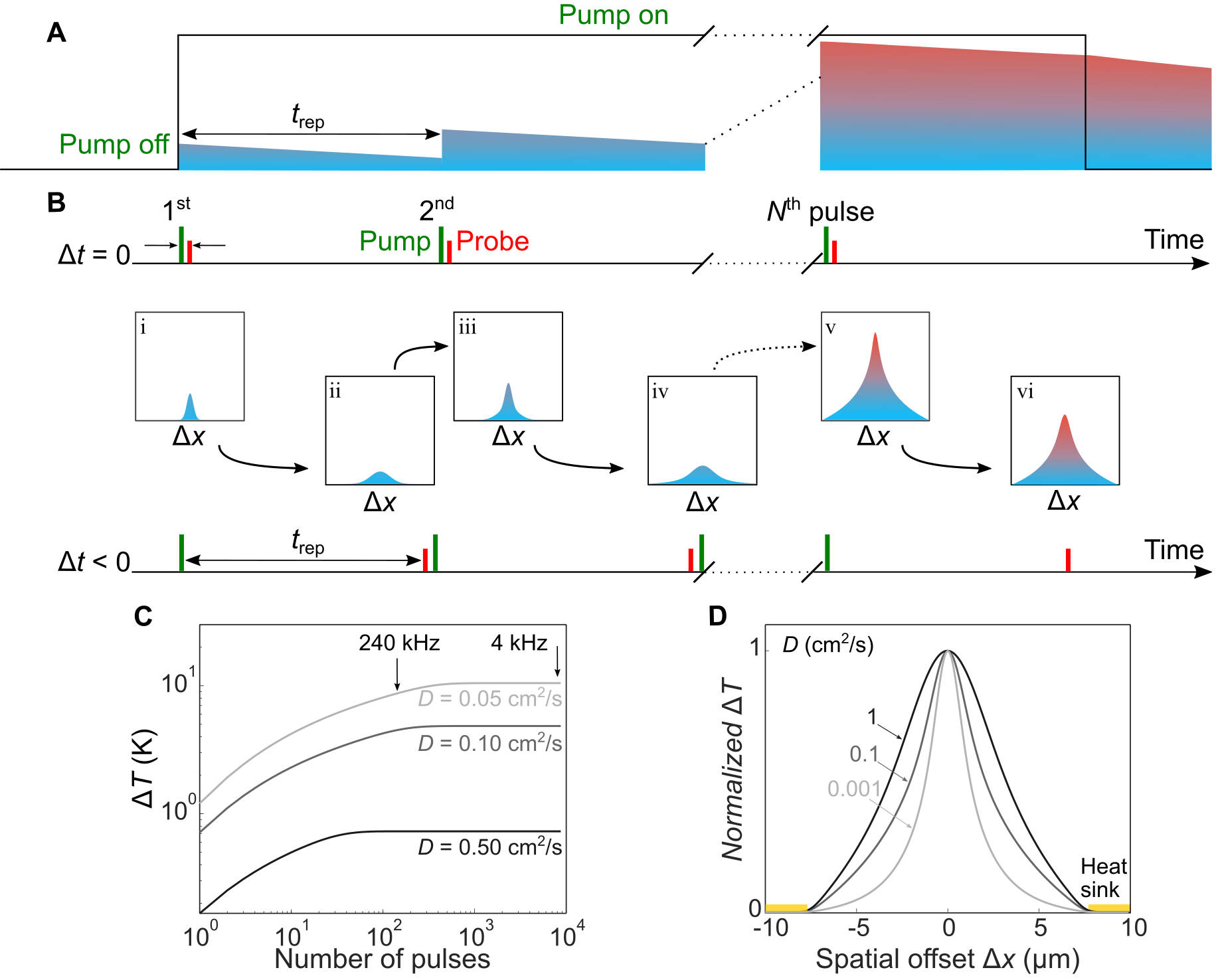}
  \caption{\textbf{Simulations of pre-time-zero profiles.} \textbf{(A)}~Heat accumulation in the free-standing region of the sample as a function of time, during one pump modulation period. \textbf{(B)}~Spatial heat profiles directly after the first pump pulse (i), after the heat of the first pump pulse has diffused (ii), directly after the second pump pulse (iii), after the combined heat of the first two pulses has diffused (iv), directly after the $N^{\rm th}$ pump pulse (v), and after the combined heat of $N$ pulses has diffused (vi). By using a negative pump-probe delay time, the $N^{th}$ probe pulse senses the effect of the $(N-1)^{th}$ pump pulse. \textbf{(C, D)}~Simulation results obtained from a straightforward Fourier heat diffusion model, as explained in the main text. \textbf{(C)}~Temperature increase $\Delta T$ as a function of the number of pulses $N$ for different input diffusivities $D$. \textbf{(D)}~Normalized average spatial heat profiles after $N$ pulses for a range of $D$. Broader average profiles correspond to higher diffusivities.} 
  \label{fgr:four}
\end{figure*}

In order to more accurately quantify the diffusivity and understand the spatial profiles, we schematically show the evolution of the temperature rise in the suspended region of the sample during one pump modulation period (Fig.~\ref{fgr:four}A). When the pump beam is not blocked, a pulse train photoexcites and locally heats the sample. Since the system does not return to equilibrium before the next pump pulse arrives, heat continuously accumulates in the system, which is excited by $N = f_{\rm rep}/2f_{\rm mod}$ pump pulses (50\% duty cycle modulation). Consequently, the system temperature increases, reaching a maximum at the end of these $N$ pump pulses. The accumulated heat then decays when the pump pulses are blocked, and the system reaches thermal equilibrium before the next period with $N$ pump pulses.

The evolution of the spatial reflectivity profiles, which correspond to temperature profiles, is shown in Fig.~\ref{fgr:four}B. After the first pump pulse, the initial photoexcitation at pump-probe delay time $\Delta t$ = 0 has a narrow Gaussian profile, see profile (i). After pump excitation and electron-phonon coupling, the heat in the phonon system starts diffusing. Therefore, at a small negative $\Delta t$, just before the next pump pulse arrives, the Gaussian profile broadens, see profile~(ii). The second pump pulse then adds a narrow Gaussian profile on top of the broadened Gaussian profile, see profile~(iii). This profile then diffuses, leading to a sum of two broadened Gaussians, see profile~(iv). This process continues for $N$ pump pulses, leading to profile~(v), just after the last pump pulse. The cumulative photoexcitation after $N$ pulses finally results in a spatial profile as in~(vi). This profile contains a narrow component that corresponds to the diffused heat from the previous pump pulse, and a broader component that corresponds to the accumulated heat from all previous pump pulses. The latter tends to the triangular steady-state solution for continuous heating and cooling through diffusion to a perfect heat sink, which is the gold-coated substrate located at 7.5 $\upmu$m from the center of the suspended region.

To understand the shape of the spatial profile quantitatively, and to obtain an accurate thermal diffusivity, we perform a simple simulation based on Fourier’s law of heat conduction:

\begin{equation*}
  \frac{\partial T}{\partial t} = D \left ( \frac{\partial^2 T}{\partial x^2} + \frac{\partial ^2 T}{\partial y^2} \right ) \hspace{1cm},
\end{equation*}
 	 
where $T$ is temperature, $t$ is time, $x$ and $y$ the spatial coordinates, and $D$ the thermal diffusivity of the material. We calculate the two-dimensional diffusion with radial symmetry, and  obtain the temperature increase $\Delta T(x,y)$. As boundary condition, we use a perfect heat sink at the edge of the circular sample with radius $r_0$. Thus, at position $r = \sqrt{x^2+y^2}$, we have $\Delta T|_{\rm r=r_{0}} = 0$~K. In the simulation, we let an initial Gaussian temperature increase evolve in space and time with a fixed diffusion coefficient. At $t$ = $1/f_{\rm rep}$, we add another pump pulse to the remnant heat profile. We repeat this for all $N$ pulses and obtain the final spatial temperature profile. For increasing $N$, we find the occurrence of saturation of the peak temperature at $\Delta x = 0$, related to the formation of a steady-state profile, see Fig.~\ref{fgr:four}C. This is in agreement with our experimental observation (Fig.~\ref{fgr:three}A) that the spatial profile does not change for the lowest modulation frequencies (larger number of pulses). The simulations indicate that with a higher diffusivity fewer pulses are needed to reach a steady-state situation, and the resulting peak temperature is lower. This is because in this case heat diffuses more quickly to the heat sink.

As a final simulation result, we show how the spatial profile changes for different diffusivities. This is important, as it indicates how sensitive our technique is towards obtaining the diffusivity of a sample, and how large a range of diffusivities can be obtained. For this, we performed simulations with different diffusion coefficients and plot the resulting average spatial profile after $N$ pulses, see Fig.~\ref{fgr:four}D. Our simulations show that the average spatial profiles get broader for larger $D$, verifying our earlier intuitive explanation. Furthermore, we see that this technique is capable of obtaining diffusivities varying over several orders of magnitude.

\subsection*{Obtaining and evaluating the thermal diffusivities of TMDs}

\begin{figure*}[htbp]
  \includegraphics[width = 13.25cm]{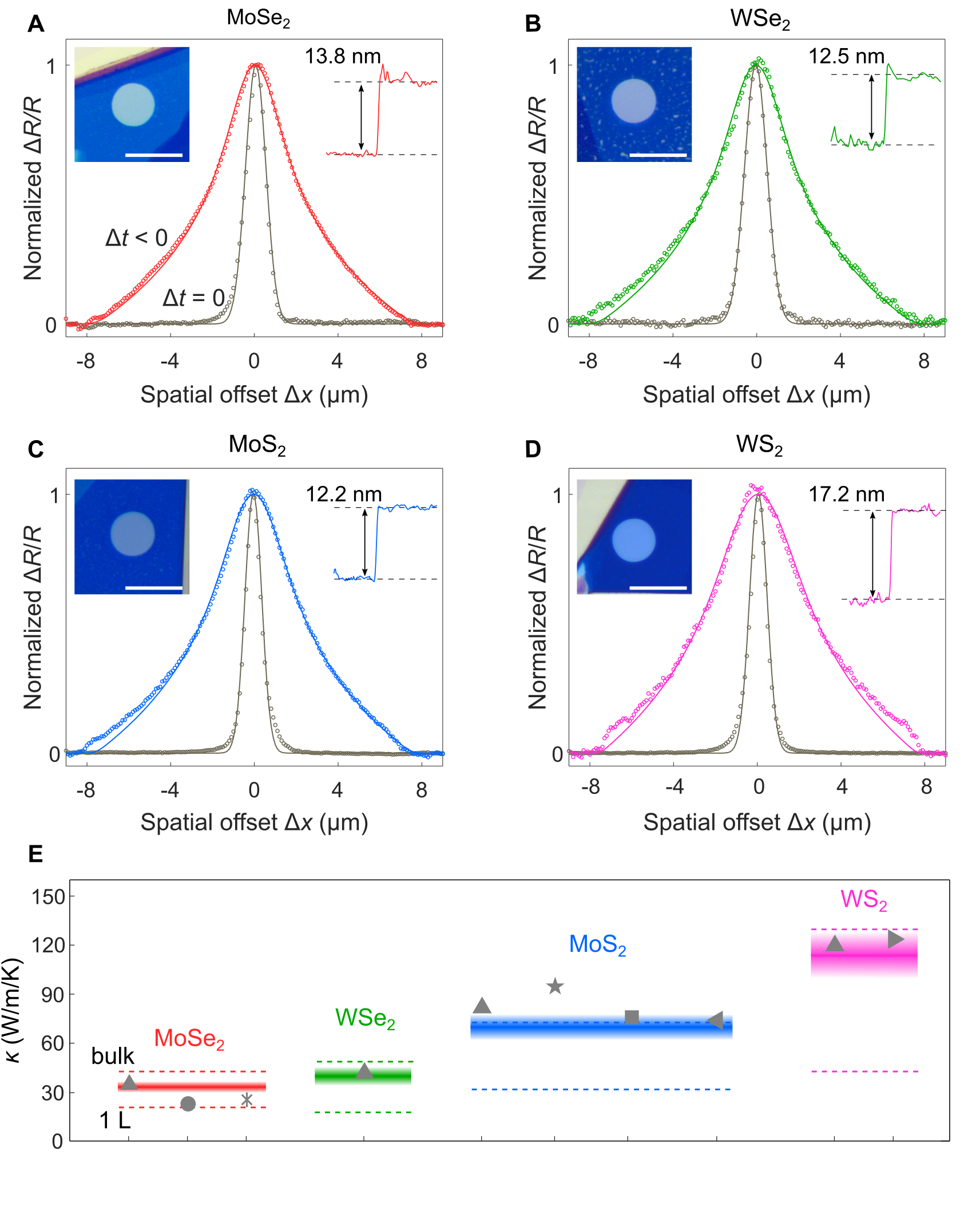}
  \caption{\textbf{Obtained thermal diffusivities.} \textbf{(A$-$D)}~Transient reflectivity $\frac{\Delta R}{R}$ profiles for suspended MoSe$_2$~\textbf{(A)}, WSe$_2$~\textbf{(B)}, MoS$_2$~\textbf{(C)} and WS$_2$~\textbf{(D)}, all normalized to the peak of the fit. The narrow profiles represent the time-zero profiles ($\Delta t$ = 0) and are described by Gaussian functions (solid grey lines) to extract their width. We obtain the thermal diffusivity $D$ by describing the pre-time-zero ($\Delta t<0$) profiles with our model based on heat diffusion (solid lines). The insets show the optical microscope image (left side, scale bar $20~\upmu$m) of the flakes, and atomic force microscopy profiles, from which we obtain the thickness (right side). \textbf{(E)}~Obtained thermal conductivity ($\kappa$) from the measured diffusivities compared with bulk experimental literature values: $\triangle$ (Ref.\cite{Jiang2017}), $\circ$ (Ref.\cite{reig2022}), $*$ (Ref.\cite{Zobeiri2019}), $\star$ (Ref.\cite{liu2014_kappa}), $\square$ (Ref.\cite{yuan2018}), $\triangleleft$ (Ref.\cite{Wang2018}), and $\triangleright$ (Ref.\cite{pisoni2016}). The thick line shows the obtained $\kappa$ for our samples with a thickness of $\sim$20 layers, where the gradient limit represents the error bound with 95\% confidence interval. The dashed lines represents our calculated $\kappa$ for bulk and monolayer (1~L).}
  \label{fgr:five}
\end{figure*}

We quantify the thermal diffusivity $D$ of four different TMD samples. We use a modulation frequency of 4.37~kHz, in order to be in the saturation regime, where the spatial profile does not depend on modulation frequency. Figure~\ref{fgr:five}A-D shows the normalized $\Delta t = 0$ and $\Delta t <$ 0 spatial profiles for MoSe$_2$~(A), WSe$_2$~(B), MoS$_2$~(C), and WS$_2$~(D), measured at a pump fluence of $\sim$0.3~J/m$^{2}$. The extracted diffusivity is the one that minimizes the squared residuals between the experimental pre-time-zero profile and the simulated profile. The only parameters we need for this are the diameter of the suspended flake ($15~\upmu$m by design), and the experimental beam width of the pump and the probe (see Supplementary Materials,~\ref{Note2}). In order to match the experimental conditions, we convolute the obtained profiles with the probe beam spatial profile. Using this procedure, we obtain thermal diffusivities $D$ of $0.18 \pm 0.02$~cm$^2$/s for MoSe$_2$, $0.20 \pm 0.03$~cm$^2$/s for WSe$_2$, $0.35 \pm 0.04$~cm$^2$/s for MoS$_2$, and $0.59 \pm 0.07$~cm$^2$/s for WS$_2$. The uncertainty in $D$ comes from two contributions: a statistical error from the  least-square fit of the data to the model (5--10\%) and a structural error mostly from the uncertainty of the  spot sizes ($\sim$5\%), see  Supplementary Materials, Fig.~\ref{fgr:SM_error} and~\ref{fgr:Freq_fit}.

In order to assess the validity of the obtained diffusivities for these TMD films, we first compare them to calculated diffusivities using density functional theory at 300 K, for both bulk crystals and monolayers by using the SIESTA program and the Temperature Dependent Effective Potential method (see Supplementary Materials,~\ref{Note3} for details). As shown recently for MoSe$_2$\cite{reig2022}, there is a modest slowdown of thermal transport towards monolayer films. We find that for all four TMDs the experimentally obtained diffusivities for our $\sim$15 nm thick films are in good agreement with calculations, and lie in between the calculated values for bulk and monolayer crystals, see Table~\ref{table:T1}. This suggests that 15 nm is a thickness where these layered materials do not yet behave completely bulk-like in terms of their thermal transport properties.

\begin{table*}[htbp]
\centering
\caption{\label{table:T1}Comparison of our experimentally obtained diffusivities for TMDs with a thickness of $\sim$20 L, with our calculated diffusivities for bulk and monolayer TMDs.}

\vspace{0.5cm}

\begin{tabular}{x{4cm} x{3cm} x{3cm} x{3cm} x{3cm}}
\toprule
\textbf{Thermal diffusivity} & \textbf{MoSe$_2$} & \textbf{WSe$_2$}  & \textbf{MoS$_2$} & \textbf{WS$_2$}  \\
 \hline
    $D_{\rm experiment, 20L}$ (cm$^2$/s) & 0.18 $\pm$ 0.02 & 0.20 $\pm$ 0.03   & 0.35 $\pm$ 0.04 & 0.59 $\pm$ 0.07 \\  \hline 
    $D_{\rm theory, bulk}$ (cm$^2$/s) & 0.24  & 0.28    & 0.39  & 0.69  \\  \hline
    $D_{\rm theory, 1L}$ (cm$^2$/s) & 0.12  & 0.10  & 0.17 & 0.23   \\ 
\bottomrule
\end{tabular}
\end{table*}

We proceed with comparing our obtained values with experimental values in the literature. Most studies report the thermal conductivity ($\kappa$), which relates to diffusivity ($D$) via the heat capacity ($C_{\rm v}$): $\kappa=D \cdot C_{\rm v}$. Therefore we convert our diffusivity values using the bulk heat capacities of MoSe$_2$ from Ref.\cite{blinder1993}, WSe$_2$ from Ref.\cite{Bolgar1990}, MoS$_2$ from Ref.\cite{volovik1978}, and WS$_2$ from Ref.\cite{ohare1984}. Figure~\ref{fgr:five}E shows the comparison of our experimental and theoretical values (solid and dashed lines respectively) to other experimental reports (symbols). We limit our comparison to literature values for bulk-like crystals of at least a few tens of nanometers. We find good agreement between our results and those from literature, confirming the general trend that the selenides (MoSe$_{2}$, WSe$_{2}$) have a lower thermal conductivity than the sulphides (MoS$_2$, WS{$_2$}).\cite{Jiang2017,lindroth2016} For bulk MoSe$_2$ and WSe$_2$, other experimental studies show thermal conductivities ranging from 22$-$42~W/m/K \cite{Jiang2017,reig2022,Zobeiri2019}, which is in excellent agreement with our obtained $\kappa$ of $33.5$~W/m/K for MoSe$_2$ and $39.6$~W/m/K for WSe$_2$ (Fig.~\ref{fgr:five}E). For WS$_2$, we find a relatively large $\kappa$ of $113.9$~W/m/K, which is in good agreement with experimental results. \cite{Jiang2017,pisoni2016} Our value for MoS$_2$ of 70 W/m/K is in good agreement with those reported in Refs.\ \cite{Jiang2017,yuan2018,Wang2018}. We do find a significantly lower thermal conductivity compared to the measured value in Ref.\ \cite{liu2014_kappa}. This discrepancy is likely the result of the smaller thickness of our sample compared to the one in Ref.\ \cite{liu2014_kappa}. Overall, we find good agreement between our experimentally obtained conductivities and the literature values. 

\subsection*{Evaluation of the pre-time-zero spatiotemporal technique}

We have seen that our pre-time-zero spatiotemporal technique allows for the accurate determination of the thermal diffusivities of thin films, and now discuss the main benefits of this technique. In comparison with  electrical  techniques, it offers the advantage of being a non-invasive, contact-free approach. Compared to other all-optical techniques, such as time-domain thermoreflectance (TDTR) \cite{Liu2013,jiang2017time,jiang2018}, frequency-domain thermoreflectance (FDTR) \cite{Schmidt2009}, techniques based on Raman thermometry \cite{Balandin2008}, and transient grating spectroscopy \cite{Kading1995},  it also has several important advantages. One of the main strengths of the technique is that it enables the determination of the in-plane thermal diffusivity of a thin film by comparing the data to a simple description based on Fourier’s heat law. Unlike most of the existing techniques, our technique does not require knowledge of any material parameters, such as the temperature coefficient, optical absorption, thickness, interfacial thermal conductance, heat capacity and/or properties related to transducer layers. The only input parameters we require for the simulations are very well known experimental conditions: the diameter of the suspended region of the material, the repetition rate of the laser, and the spot sizes of the pump and probe beams. Since all these values are known within an error of a few percent or less, we obtain highly accurate diffusivities (see Supplementary Materials, Fig.~\ref{fgr:SM_error}). Thus the proposed method eliminates sources of error arising from necessary measurements of intermediate parameters that enter the model used for the extraction of unknown thermal properties. 

Another important advantage is the sensitivity of the technique. The ability to observe changes in a spatial profile is limited only by the signal-to-noise ratio, which ultimately determines the accuracy in determining the diffusivity. Being a pump-probe technique, where pump-induced changes are detected differentially using the combination of an optical chopper and a synchronized lock-in amplifier (Zurich Instruments MFLI), means that we can resolve very subtle heat-induced changes in permittivity (refractive index) of materials. As a result, it is possible to determine thermal transport properties even if the change in temperature due to heating is very small, on the single Kelvin level (see Fig.~\ref{fgr:two}E and \ref{fgr:two}F). This is an important advantage over several alternative techniques that typically require stronger heating.
 
Beyond offering these advantages, pre-time-zero spatiotemporal pump-probe microscopy will make it possible to explore previously inaccessible physical phenomena and properties. Continuing with thermal transport as the main example, we suggest combining our technique that provides the diffusivity with a complementary study that provides the thermal conductivity, in order to obtain the heat capacity in an all-optical fashion. Our technique will also likely make it possible to access transport regimes where diffusive transport breaks down, and ballistic or hydrodynamic phonon transport occurs. These phenomena have been studied theoretically for two-dimensional materials \cite{cepellotti2015,lee2015}, however there is minimal experimental evidence to support these reports \cite{xu2014,block2021}. As our technique is highly sensitive to diffusive processes that occur over nanometer length scales, we predict that it will facilitate an advance in the understanding of non-diffusive transport in low-dimensional materials. We furthermore point out that our technique enables the study of more complex structures, such as van der Waals heterostructures consisting of combinations of different layered materials, with material specificity through the material-specific exciton resonances. 
 
Finally, we note that the concept of pre-time-zero spatiotemporal microscopy is generally applicable to a broad range of material systems. In our example of thermal diffusion in thin layered films, we benefit from an increased sensitivity due to the temperature-dependent exciton response. However, the technique will likely work for any diffusing species that has an optical response associated with it and that has a long enough lifetime to be detectable at negative time delay. Since temperature-dependent refractive indices are very common, we expect that the thermal diffusivity of materials without an exciton response should also be attainable. We note that the thermal diffusivity might not be directly obtainable in all materials. For example, the pre-time zero profile for materials with very long-lived photoexcited charges, such as halide perovskites, will contain a significant contribution that corresponds to electronic diffusion, rather than thermal diffusion. This, however, means that this technique makes it possible to observe charge diffusion in semiconducting thin films with nanosecond or microsecond carrier lifetimes, which is particularly interesting for such materials with low electronic diffusivities. 

\section*{CONCLUSIONS}

In this work, we have introduced a technique for measuring diffusion processes of long-lived excitations by exploiting the pre-time-zero spatial profile. In particular, we have shown that this technique can be used to obtain the in-plane thermal diffusivity of thin films. We have determined the thermal diffusivities of suspended MoSe$_2$, WSe$_2$, MoS$_2$, and WS$_2$ thin films. Our technique provides a simple, sensitive, and accurate way of determining the thermal diffusivity, and will enable a deeper understanding of the thermal properties of nanoscale materials. Furthermore, the technique of pre-time-zero spatiotemporal scanning constitutes a powerful tool for studying a multitude of diffusion processes in a large variety of systems.

\section*{ACKNOWLEDGEMENTS} 
ICN2 was supported by the Severo Ochoa program from Spanish MINECO Grant No. SEV-2017-0706. S.V. and D.S.R. acknowledge  the  support  of  the  Spanish  Ministry of Economy through FPI-SO2018 and FPI-SO2019, respectively. K.J.T. acknowledges funding from the European Union's Horizon 2020 research and innovation program under Grant Agreement No. 804349 (ERC StG CUHL), RYC fellowship No. RYC-2017-22330 and IAE project PID2019-111673GB-I00. ICFO was supported by the Severo Ochoa program for Centers of Excellence in R$\&$D (CEX2019-000910-S), Fundació Privada Cellex, Fundació Privada Mir-Puig and the Generalitat de Catalunya through the CERCA program. P.W. acknowledges funding from the European Union's Horizon 2020 research and innovation programme under the Marie Skłodowska-Curie grant agreement No. 754510 (PROBIST). N.F.v.H. acknowledges funding by the European Commission (ERC AdG 670949-LightNet), the Spanish Plan Nacional (PGC2018-096875-BI00) and the Catalan AGAUR (2017SGR1369). M.J.V. acknowledges support from Fédération Wallonie Bruxelles and ULiège (ARC project DREAMS G.A. 21/25-11). Z.Z. acknowledges the research program “Materials for the Quantum Age” (QuMAt) for financial support. This program (registration number 024.005.006) is part of the Gravitation program financed by the Dutch Ministry of Education, Culture and Science (OCW). We acknowledge PRACE computing time on MareNostrum4 at Barcelona Supercomputing Center (OptoSpin id. 2020225411). \vspace{0.25cm}

“The following article has been submitted to \textit{Review of Scientific Instruments}. After it is published, it will be found at~\href{https://publishing.aip.org/resources/librarians/products/journals/}{Link}.” 

\bibliography{references.bib}


\clearpage

\newpage
\onecolumngrid

\SupplementaryMaterials

\section*{Supplementary Materials}
\vspace{1.5cm}
\begin{figure}[ht!]
    \centering
      \includegraphics[width = 13cm]{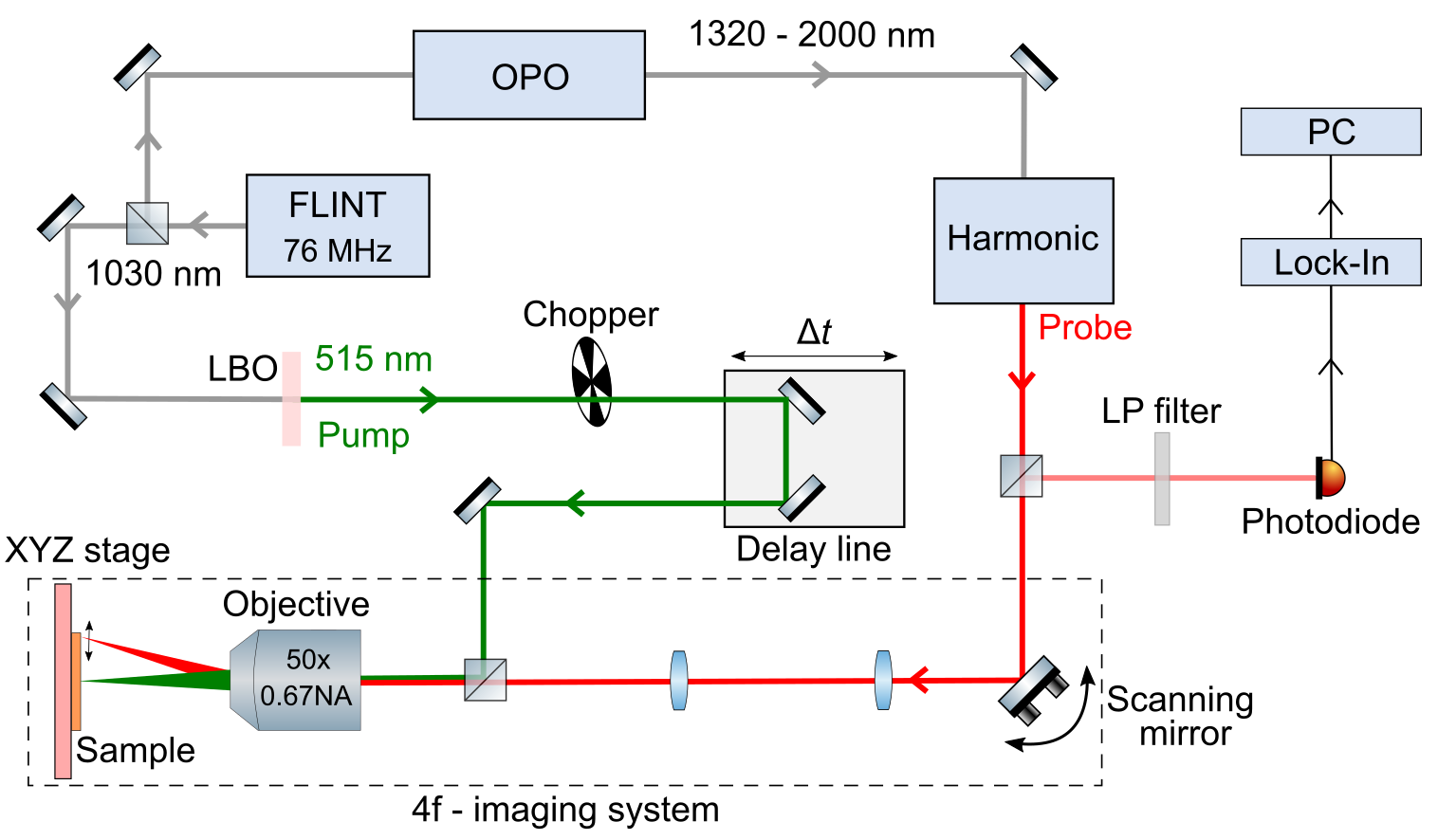}
      \caption{\textbf{Detailed schematic of  the  pre-time-zero spatiotemporal setup.} LBO:~Lithium triborate crystal. LP:~longpass filter. Description of the experiment is found in the main text.}
      \label{fig:SM_setup}
\end{figure}

\clearpage

\begin{figure}[!ht]
  \includegraphics[width = 12cm]{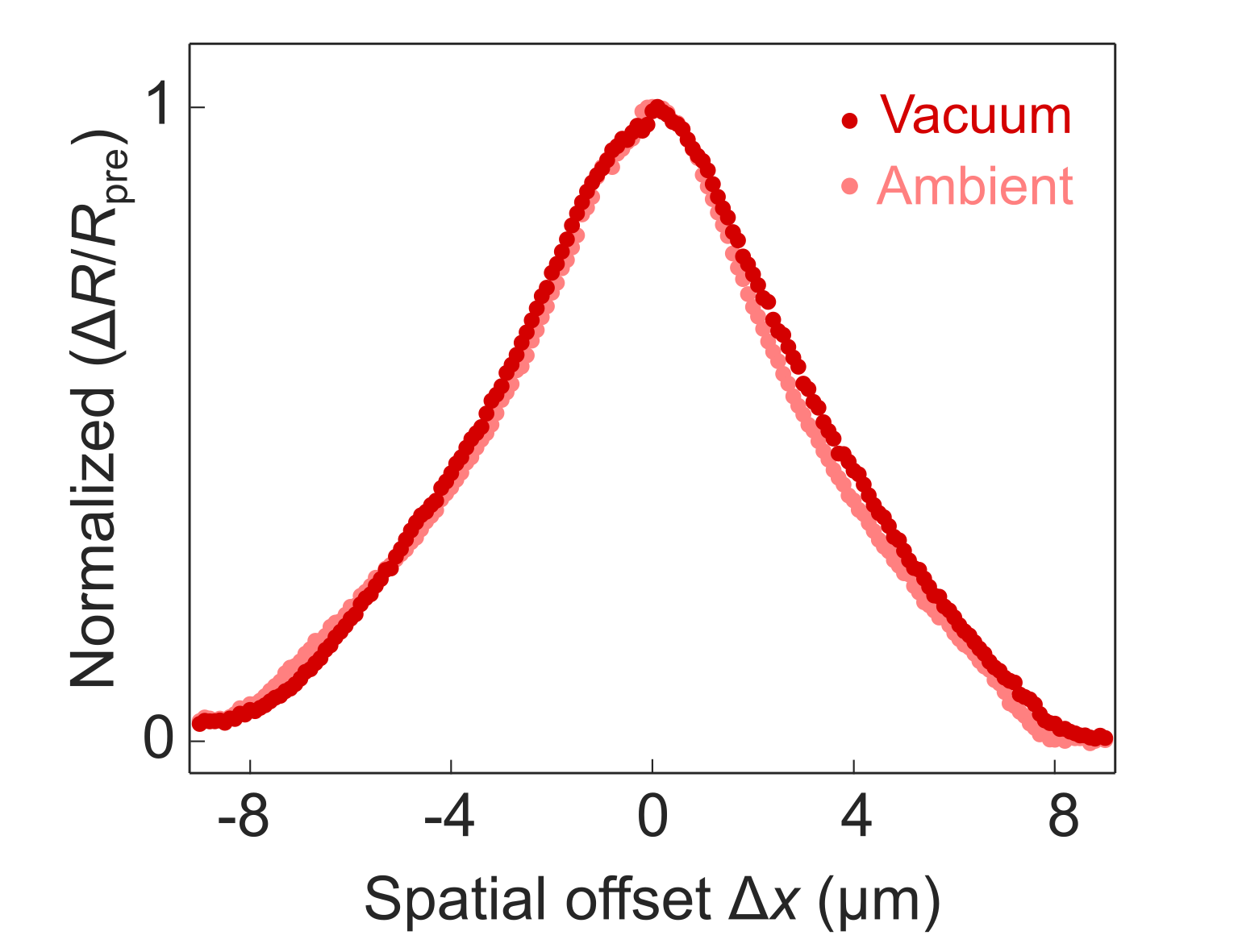}
  \caption{\textbf{Ambient vs vacuum measurements.} Pre-time-zero ($\frac{\Delta R}{R}_{\rm pre})$ spatial profiles on MoSe$_2$ in ambient and vacuum conditions. The difference between vacuum and air environment is very small at this thickness (14~nm) range, as the dominant heat transport mechanism is the in-plane diffusion towards the heat sink. Profiles are normalized to their peak value.} 
  \label{fgr:SM_vacuum}
\end{figure}

\clearpage

\begin{figure}[! ht] 
  \includegraphics[width = 12cm]{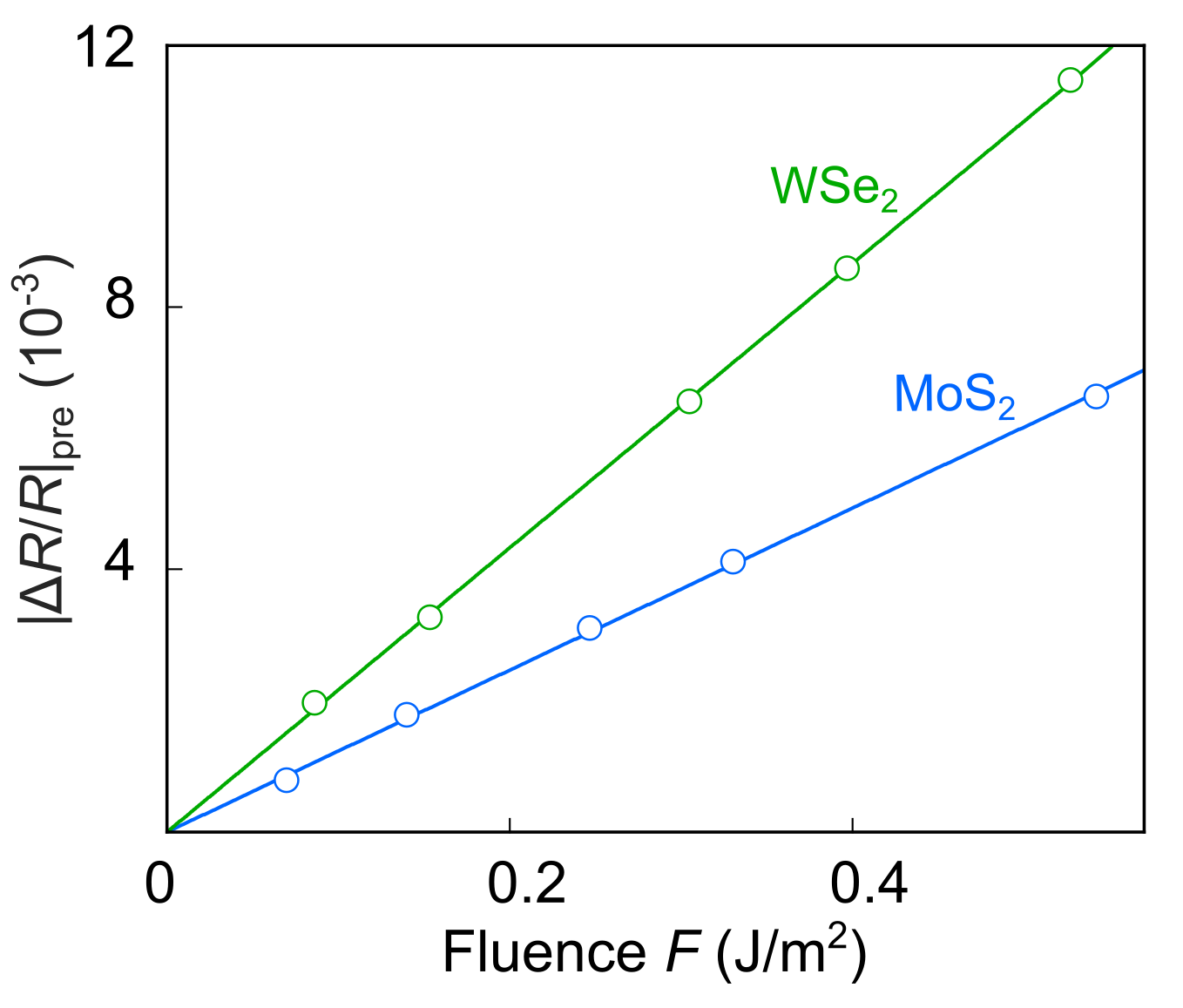}
  \caption{\textbf{Pre-time-zero vs.~fluence.} Linear dependence of |$\frac{\Delta R}{R}|_{\rm pre}$  signal with incident fluence on WSe$_2$ and MoS$_2$.} 
  \label{fgr:SM_fluence}
\end{figure}

\clearpage

\subsection{\label{Note1} Lorentz Oscillator Model}
We describe the complex dielectric function of WS$_2$ and MoSe$_2$ as a function of angular frequency $\omega$ using the Lorentz oscillator model for the complex permittivity $\epsilon$:

\begin{equation*}
  \epsilon (\omega) = (n_{\rm b} + ik_{\rm b})^2  + \frac{A}{(\omega_{\rm 0} - \omega^2 - i\gamma \omega)}
   \hspace{1cm},
\end{equation*}

where $n_{\rm b} + ik_{\rm b}$ represents the background complex refractive index in the absence of excitons and was assumed to be equal to that of bulk material \cite{beal1976,hsu2019,beal1979}. The exciton resonance energy is denoted by $\omega_{\rm 0}$ with amplitude $A$, and a phenomenological (temperature-dependent) broadening parameter $\gamma$ equal to the full-width at half maximum of a Lorentzian function. We calculated the dielectric response for various exciton linewidths starting from room temperature. The complex refractive index is then obtained from $n=\sqrt{\epsilon(\omega)}$. Using the thin film Fresnel equation, we plot the reflection ($R$) profile for different linewidths (and therefore temperatures). The transient reflection is calculated using $\frac{\Delta R}{R}$ = $\frac{R_{\rm (T)} - R_{(\rm T = 300~K)}}{R_{(\rm T = 300~K)}}$, where $R_{\rm (T = 300~K)}$ is the reflectivity at room temperature.
  

\begin{figure} [! ht] 
  \includegraphics[width = 12cm]{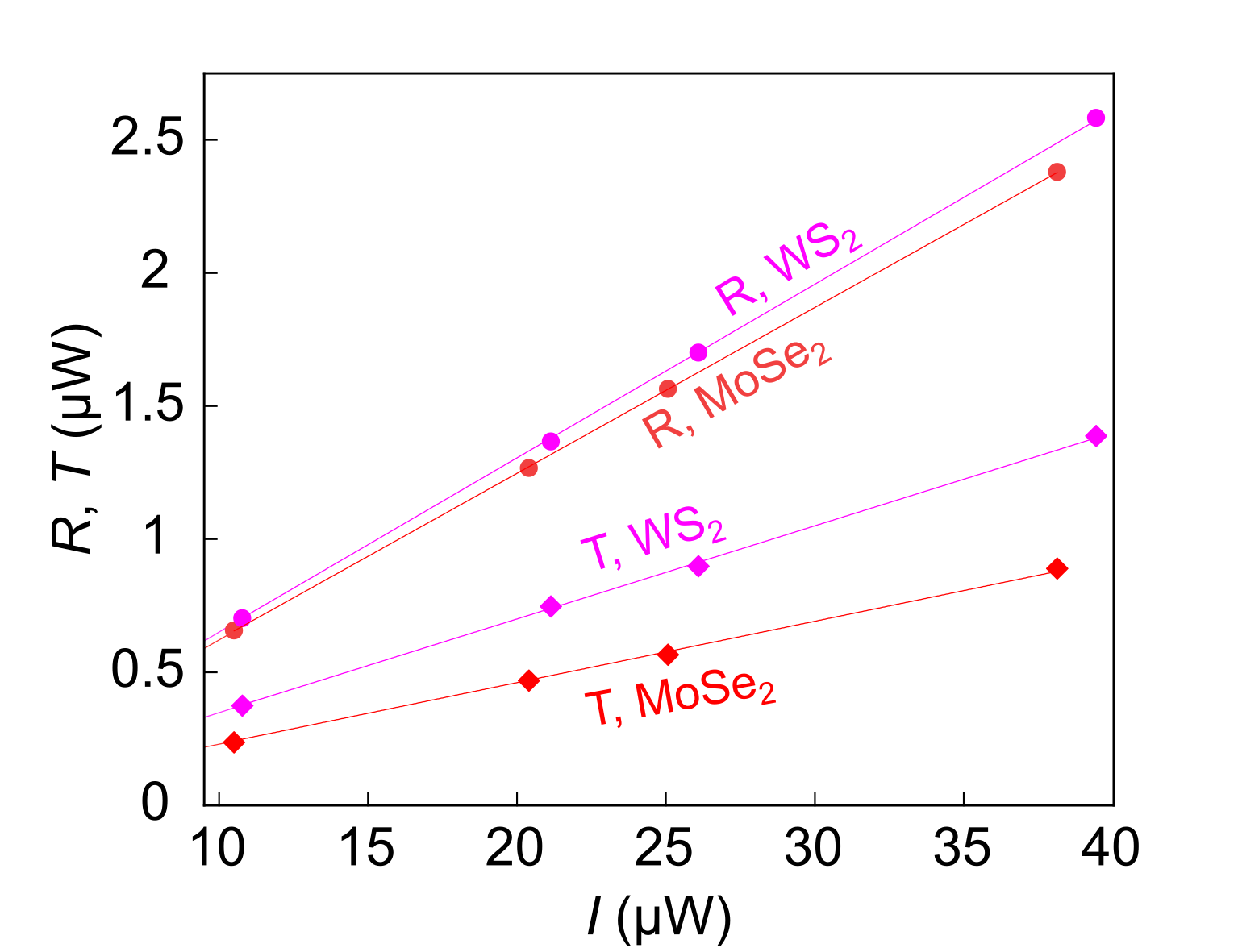}
  \caption{\textbf{Absorption measurements at 515~nm.} Reflected ($R$) and transmitted ($T$) power (scatter) from suspended MoSe$_2$ and WS$_2$ flakes with Incident ($I$) power. The solid line is the linear fit to the data. The slope of these fits are used to determine the absorption ($A$) of the flakes, using the following relation: $A$ = $[1 - $ $\frac{T_{\rm sample}}{T_{\rm 0}}$ $- (\frac{R_{\rm sample} - R_{\rm 0}}{R_{\rm 100}})] \times 100$. Here ${T_{\rm sample}}$ and ${R_{\rm sample}}$ is the slope of $I - T$ and $I - R$ plot of respective samples, $T_{\rm 0}$ and $R_{\rm 0}$ is the slope of $I - T$ and $I - R$ plot respectively, of an empty 15 micron hole substrate, and $R_{\rm 100}$ is the slope of $I - R$ plot of a perfect mirror. We obtained $A$ of 33.2\% and 23.9\% for MoSe$_2$ and WS$_2$, respectively.}
  \label{fgr:SM_absorption}
\end{figure}

\clearpage

\begin{figure}[! ht] 
  \includegraphics[width = 12cm]{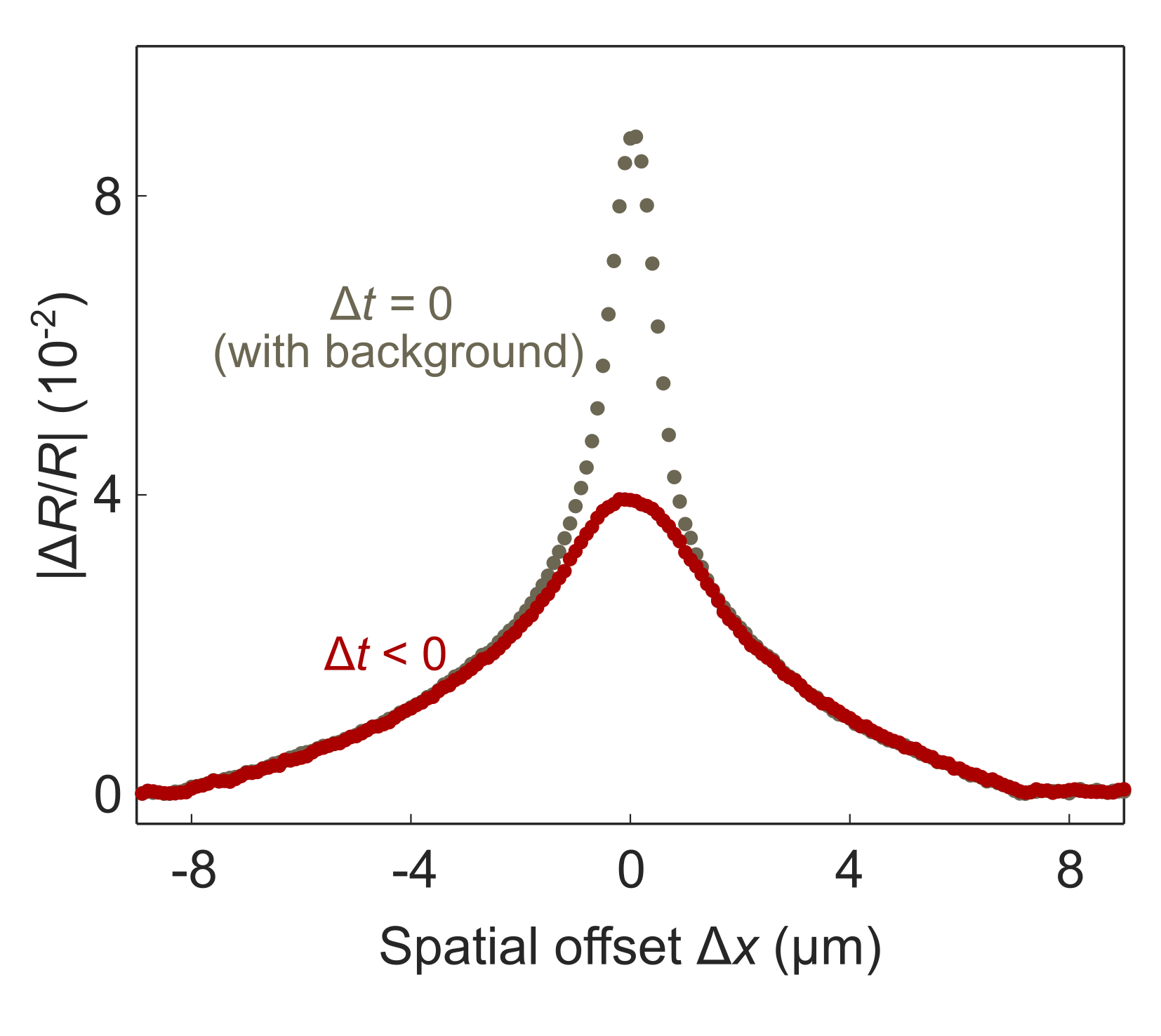}
  \caption{\textbf{Spatial profiles and background removal.} Exemplary spatial profiles (MoSe$_2$) showing the $\Delta t = 0$ (with background, grey circles) and $\Delta t < 0$ profile. We remove the background signal ($\Delta t < 0$) to obtain the corresponding time-zero ($\Delta t = 0$) profile.} 
  \label{fgr:SM_background}
\end{figure}

\clearpage

\begin{figure}
  \includegraphics[width = \linewidth]{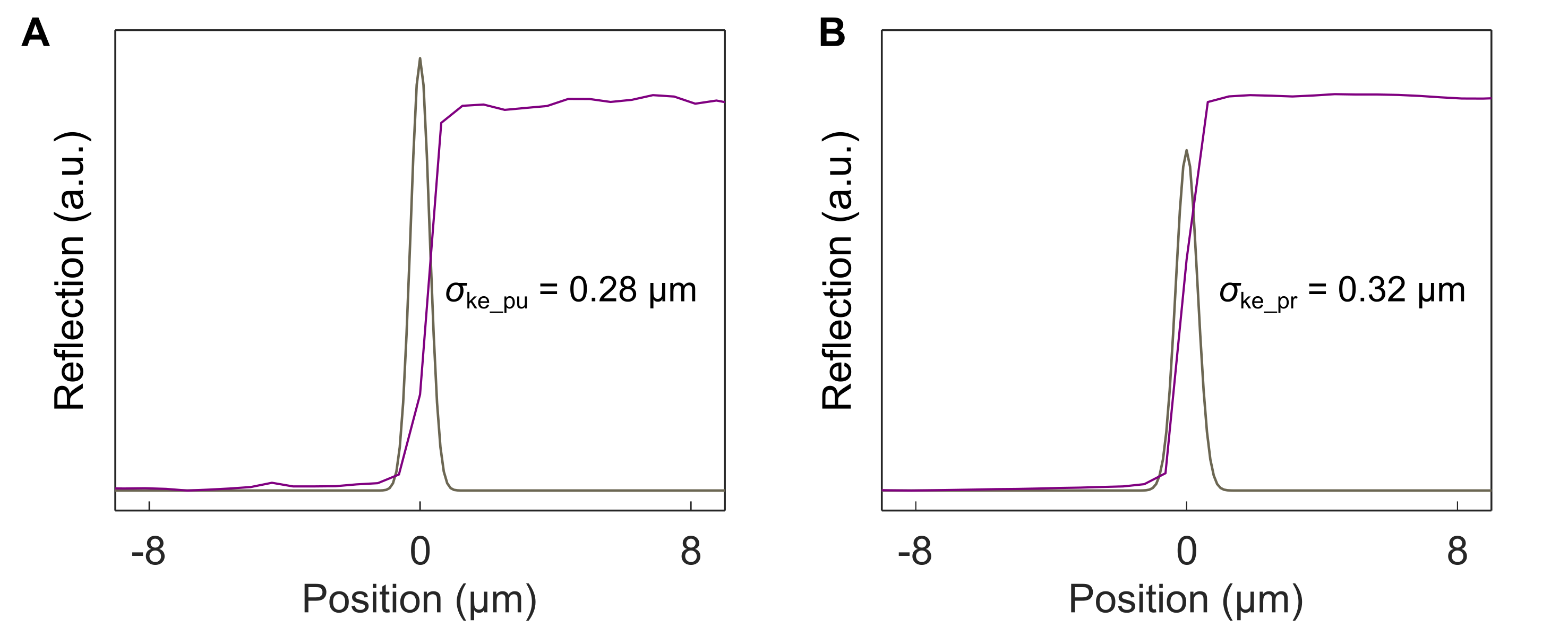}
  \caption{\textbf{Spot size measurement.} An exemplary knife edge measurement for pump~\textbf{(A)} and probe~\textbf{(B)} performed at a sharp metal (Au) edge, while tracking the reflection signal (purple line). From the Gaussian fit to its derivative (grey line) we extract the width of pump ($\sigma_{\rm ke\_pu}$) and probe beams ($\sigma_{\rm ke\_pr}$).}
  \label{fgr:SM_knifeedge}
\end{figure}

\subsection{\label{Note2} Experimental Beam Width Calculation}
Prior to our spatiotemporal measurements we characterize the beams by measuring the beam profile with a knife edge method \cite{suzaki1975}, and determine the width of the pump ($\sigma_{\rm ke\_pu}$) and probe beams ($\sigma_{\rm ke\_pr}$). To calculate the individual experimental beam widths ($\sigma_{\rm pu}$ and $\sigma_{\rm pr}$), first we fit the time-zero spatial profile (after subtracting $\Delta t < 0$) with a single Gaussian function and obtain the width ($\sigma_{\Delta t = 0}$). As mentioned before, this shape corresponds to the convoluted pump and the probe profiles. Hence, we solve the following relation
\begin{equation*}
  \sigma^2_{\rm pu} + \sigma^2_{\rm pr} = \sigma^2 _{\Delta t = 0}
  \hspace{1cm},
\end{equation*}
to obtain the experimental beam widths by assuming the ratio between experimental beam widths remains identical to the knife-edge measurements ($\sigma_{\rm ke\_pu}$/$\sigma_{\rm ke\_pr}$ = $\sigma_{\rm pu}$/$\sigma_{\rm pr}$).
 

\subsection{\label{Note3} Density  Functional  Theory  Simulations}
We employed density functional theory (DFT) calculations as implemented in the SIESTA software package \cite{garcia2020}. For the simulations we employed Norm-Conserving (PBE) Pseudopodentials from the PseudoDojo Library generated using the ONCVPSP software \cite{hamann2013} and the GGA-PBE exchange-correlation functional without spin-orbit coupling \cite{perdew1996}. To take into account the long-range electron-electron correlation, we use the vdW-DF2 LMKLL functionals \cite{lee2010}. Calculations follow energy and k-point mesh convergence studies with a 1000 Ry energy cutoff for the real-space grid with a (20 $\times$ 20 $\times$ 1) \textbf{k}-points sampling of the Brillouin zone for the monolayers and (20 $\times$ 20 $\times$ 20) \textbf{k}-points grid for the bulk crystals. A standard double zeta polarized (DZP) basis and an electronic temperature of 300~K was used. For the relaxation of the structural parameters we employed the conjugate gradient algorithm until the forces on the atoms became smaller than 0.001~eV\AA$^{-1}$ and the maximum stress component was smaller than 0.5~GPa. In the monolayer case we applied 17~\AA~of vacuum in order to avoid interactions between two periodical images. Atomic forces are computed by using a supercell approach (10 $\times$ 10 $\times$ 1 for monolayers and 8 $\times$ 8 $\times$ 2 for bulk crystals) and the thermal properties are then computed with TDEP \cite{hellman2011} by iteratively solving the full Boltzmann transport equation. With this method we can sample a canonical ensemble at a given temperature by generating supercells with displaced atoms according to the Bose-Einstein statistics. From them, by using DFT, we can extract the atomic forces and fit them in order to obtain the harmonic and anharmonic interatomic forceconstants, needed for the calculation of the heat capacity and the lattice thermal conductivity. 
 

\begin{figure}[! ht] 
  \includegraphics[width = 12cm]{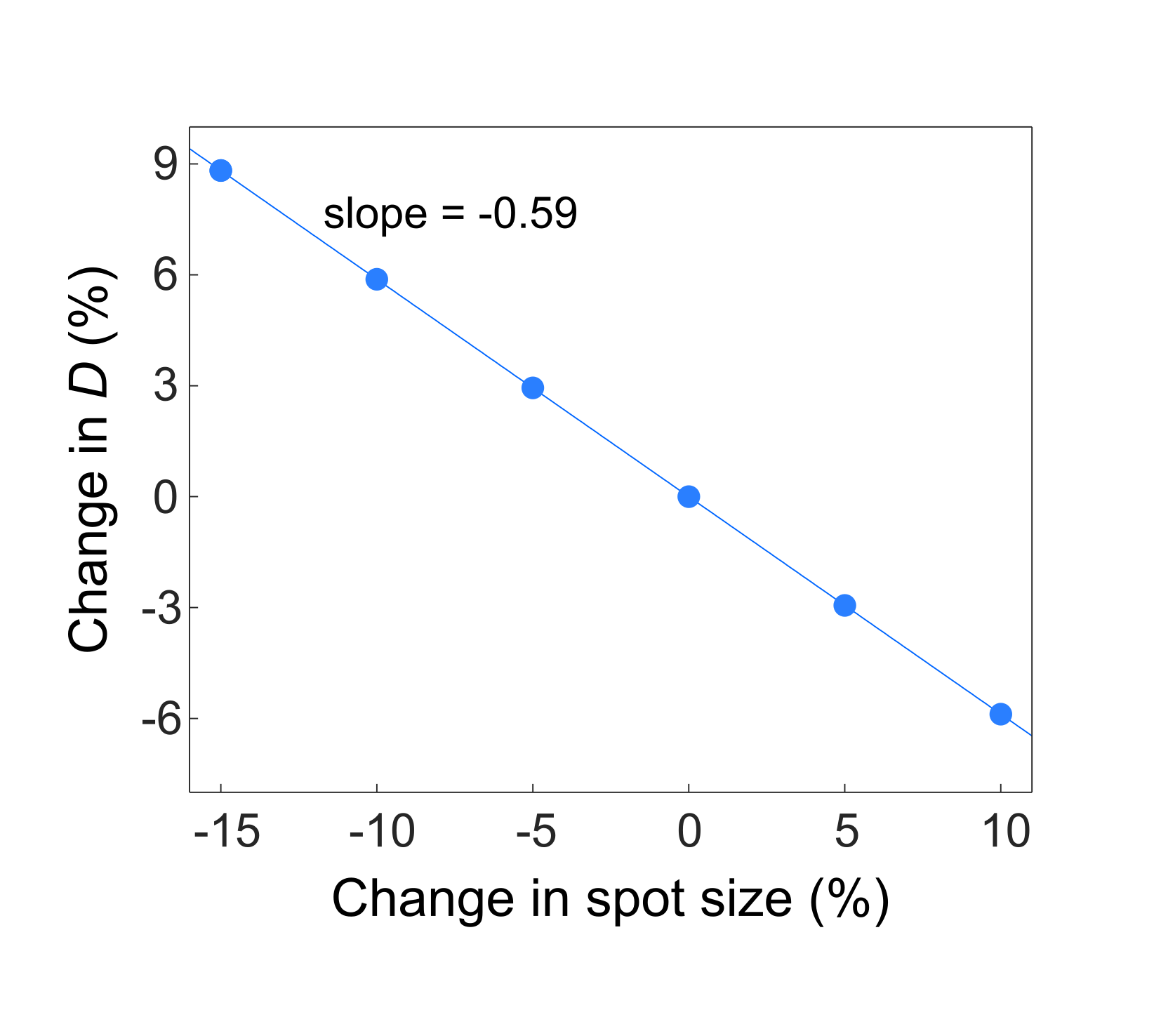}
  \caption{\textbf{Effect of spot size in quantifying thermal diffusivity.} Here, we varied the width of the pump beam (heat source) 
  in simulation, and described the respective simulated profiles with $\Delta t < 0$ profile of our MoSe$_2$ sample. A change in 10\% of pump spot size changes the extracted $D$ by $\approx$ 6\%.} 
  \label{fgr:SM_error}
\end{figure}
\clearpage

\begin{figure}[! ht] 
  \includegraphics[width = \linewidth]{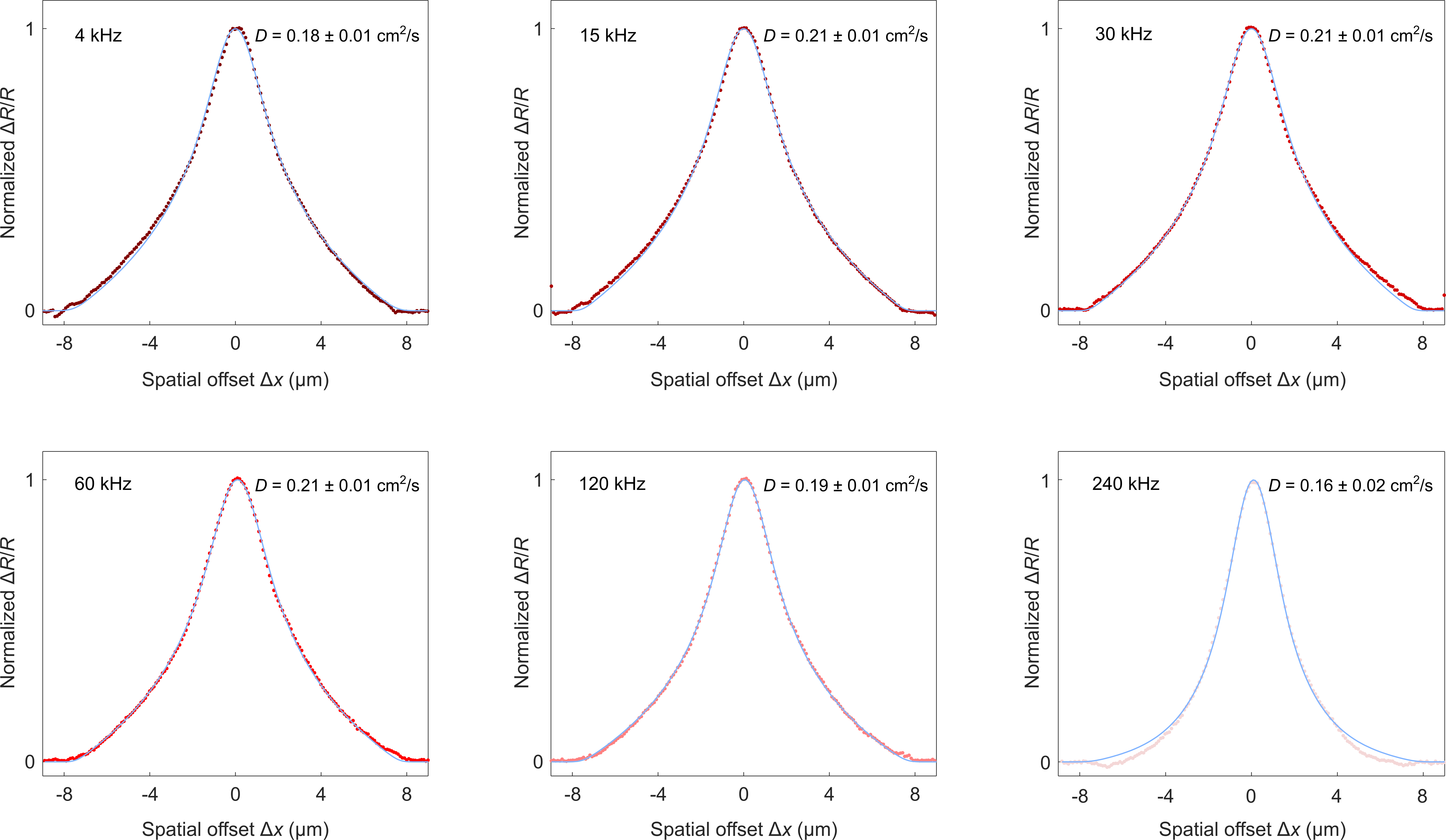}
  \caption{\textbf{Diffusivities at different modulation frequencies}. Diffusivity determination of MoSe$_2$ for different modulation frequencies, providing an estimate of the error in determining $D$. The solid circles are the experimental data points fitted using our heat diffusion model (solid lines).} 
  \label{fgr:Freq_fit}
\end{figure}
\clearpage

\begin{figure}[! ht] 
  \includegraphics[width = 12cm]{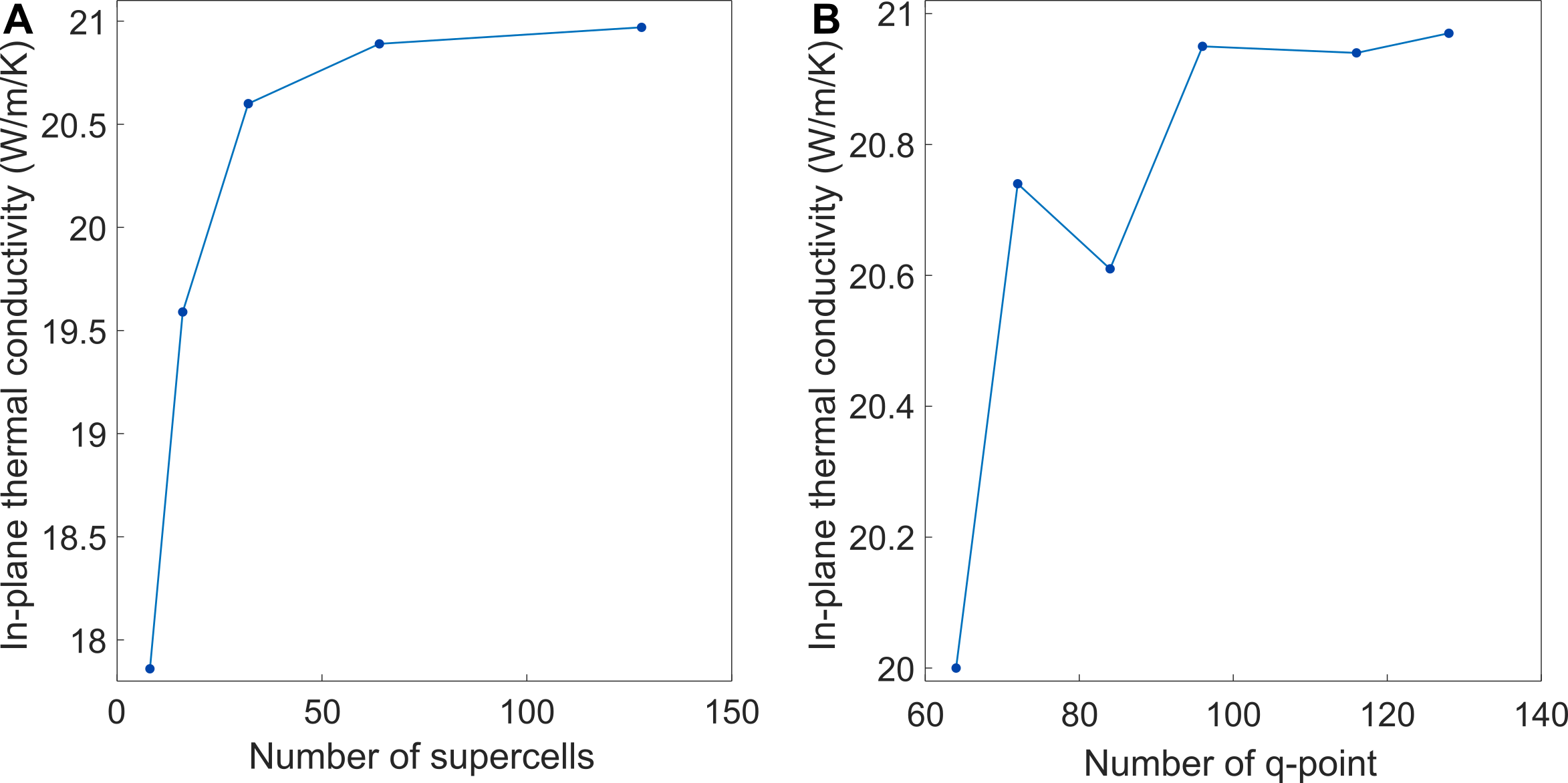}
  \caption{\textbf{Convergence tests for Density Functional Theory simulations}. \textbf{(A)} We checked the number of supercells used for the calculations of forces by looking at the thermal conductivity change. We considered converged the calculation when the value was changing by less than 1\%. \textbf{(B)} We checked  the number of q-points used for integrating the phonon grid. Here, q-points along the direction x = 20, y = 20, and z = 1 (20) for monolayer (bulk). In our calculation we perform thermal conductivity calculations until the value changed by less than 1\%. The tests reported here refer to monolayer MoSe$_2$.}
  \label{fgr:Convergence}
\end{figure}
\clearpage

\begin{figure}[! ht] 
  \includegraphics[width = 10cm]{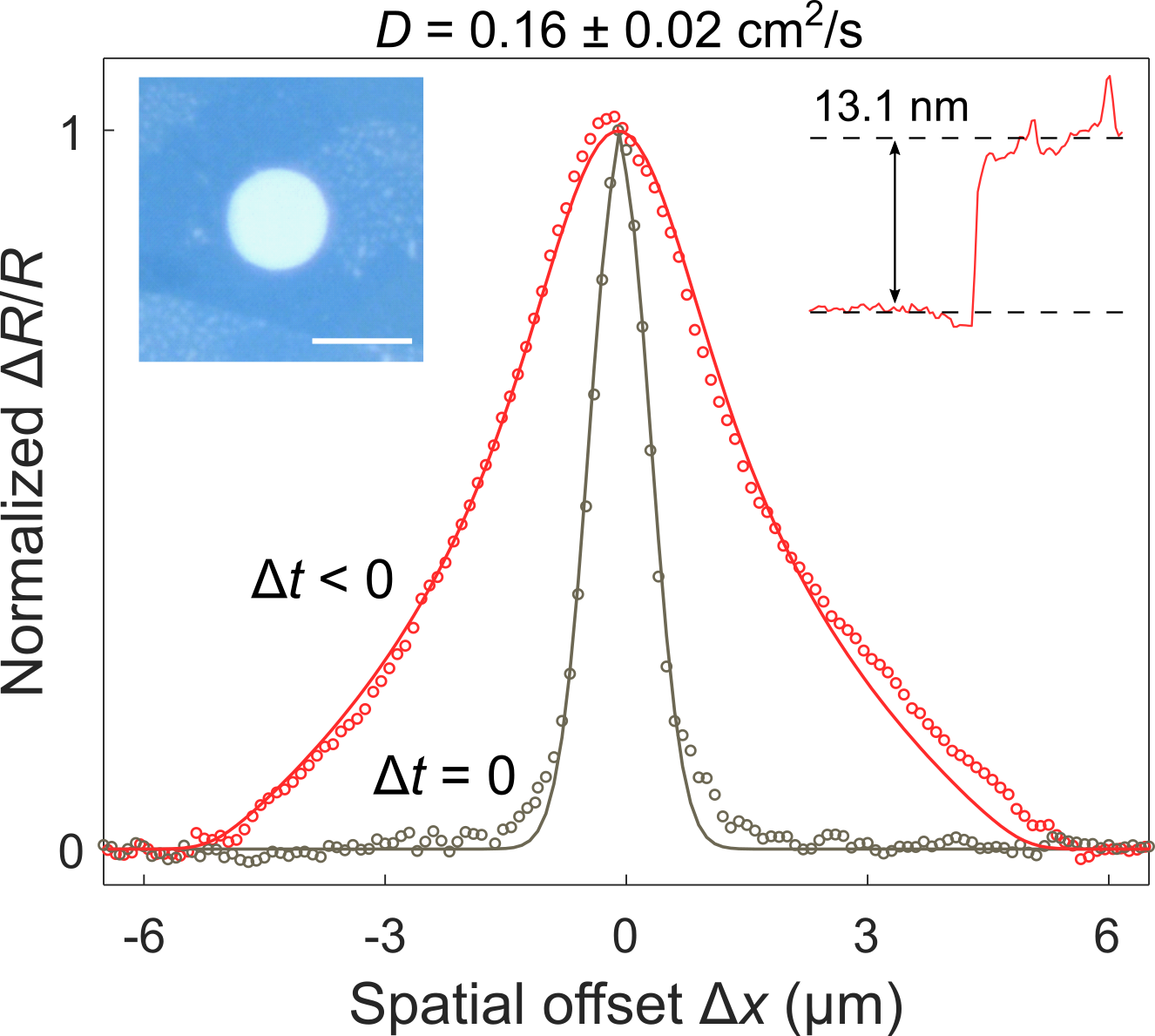}
  \caption{\textbf{Measurements on \textbf{$10~\upmu$m} suspended sample}. Transient reflectivity $\frac{\Delta R}{R}$ profiles for an MoSe$_2$ sample suspended over a  hole with a diameter of $10~\upmu$m, normalized to the peak of the fit. The narrow profile represent the time-zero profile ($\Delta t$ = 0) and is described by Gaussian function (solid grey line) to extract the width. The solid red line is the fit to the pre-time-zero ($\Delta t<0$) profile. The insets show the optical microscope image (left side, scale bar $10~\upmu$m) of the flakes, and atomic force microscopy profiles, from which we obtain the thickness (right side). The obtained diffusivity for this sample agrees with the results obtained for the flake suspended over a hole with a diameter of $15~\upmu$m, adding credibility to the validity of our technique. } 
  \label{fgr:Exp_10um}
\end{figure}

\end{document}